\documentclass[letterpaper,9pt]{article}
\usepackage{authblk}
\usepackage{tabularx} 
\usepackage{amsmath}  
\usepackage{graphicx} 
\usepackage[margin=1in,letterpaper]{geometry} 
\usepackage[final]{hyperref} 
\usepackage[sorting=none, url=false, doi = false]{biblatex}
\usepackage{dirtytalk}
\hypersetup{
	colorlinks=true,       
	linkcolor=blue,        
	citecolor=blue,        
	filecolor=magenta,     
	urlcolor=blue         
}

\def\fnl{f_{NL}}

\def\x{{\bf x}}
\def\k{{\bf k}}

\title{A Tale of Two Fields: Neural Network-Enhanced non-Gaussianity Search with Halos}

\author[1]{Yurii Kvasiuk} 
\author[1]{Moritz M\"unchmeyer} 
\author[2]{Kendrick Smith}

\affil[1]{Department of Physics, University of Wisconsin-Madison, Madison, WI 53706, USA}
\affil[2]{Perimeter Institute of Theoretical Physics, Waterloo, ON N2L 2Y5, CA}

\begin{document} 
\include{preamble}

\maketitle

\begin{abstract}
    It was recently shown that neural networks can be combined with the analytic method of scale-dependent bias to obtain a measurement of local primordial non-Gaussianity, which is optimal in the squeezed limit that dominates the signal-to-noise. The method is robust to non-linear physics, but also inherits the statistical precision offered by neural networks applied to very non-linear scales. In prior work, we assumed that the neural network has access to the full matter distribution. In this work, we apply our method to halos. We first describe a novel two-field formalism that is optimal even when the matter distribution is not observed. We show that any N halo fields can be compressed to two fields without losing information, and obtain optimal loss functions to learn these fields. We then apply the method to high-resolution AbacusSummit and AbacusPNG simulations. In the present work, the two neural networks observe the local population statistics, in particular the halo mass and concentration distribution in a patch of the sky. While the traditional mass-binned halo analysis is optimal in practice without further halo properties on AbacusPNG, our novel formalism easily allows to include additional halo properties such as the halo concentration, which can improve $\fnl$ constraints by a factor of a few. We also explore whether shot noise can be lowered with machine learning compared to a traditional reconstruction, finding no improvement for our simulation parameters. 
\end{abstract}

\section{Introduction}

The most precise constraints on cosmology so far are given by the CMB data. However, the modern and near-future Large Scale Structure Surveys are expected to reach and surpass the level of sensitivity of CMB experiments, especially to one of the key probes of the inflationary physics - the amplitude of the local primordial non-Gaussianity - $f^{loc}_{NL}$ \cite{Alvarez:2014vva}. Several ways to measure $f^{loc}_{NL}$ from the LSS data have been proposed so far, such as field-level forward-modelling \cite{Andrews_2023}, topological methods, e.g. \cite{Biagetti_2021}, estimation from the squeezed bispectrum \cite{Goldstein_2022}, etc. Measurements of the several probes that trace the same underlying matter field can be combined to leverage the cancellation of the sample variance, resulting in enhanced sensitivity \cite{Seljak_2009, Castorina_2018}. Such probes can be derived from CMB secondary anisotropies, such as lensing \cite{Schmittfull_2018} or kSZ effect \cite{M_nchmeyer_2019, Giri_2022, smith2018ksztomographybispectrum, Cayuso_2023, Deutsch_2018, Contreras_2023, kvasiuk2024autodiff}. This work continues to develop a neural-network-based approach.
When applied to the non-linear regime of structure formation, neural networks can in principle extract significantly more information about cosmological parameters than conventional probes such as the power spectrum. Such measurements are not in generally robust to unknown non-linear physics, although there are significant simulation efforts to explore (and ultimately marginalize over) different non-linear models. 

However, \cite{Giri_2023} showed that the situation is much better in the case of constraining local non-Gaussianity $\fnl$, if the neural network is used in a specific way. A classic result in cosmology shows that $\fnl$ introduces scale-dependent bias in the halo distribution. In \cite{Giri_2023} it was argued that any measurement of the local amplitude of structure $\sigma_8^{loc}$ will have scale-dependent bias. In particular, a neural network can be trained to measure $\sigma_8^{loc}$ across the sky. If the neural network can measure $\sigma_8^{loc}$ better than the local halo density, it will provide a field that is more sensitive to $\fnl$ than traditional scale-dependent bias. In \cite{Giri_2023}, the experimentally unrealistic simplifying assumption was made that the matter distribution is exactly known and that the neural network has access to it and in \cite{Giri:2023mpg}, the same approach is implemented with local power spectrum as a summary statistic and tested on both non-linear matter field and halo catalogs. 
Our main goal in this paper is to proceed with a neural-network-based approach but use only a simulated halo catalogue.

In the first part of this paper, we generalize the work of \cite{Giri_2023} from a single neural network field $\pi^{NN}$ to two fields $\pi_m^{NN}$ and $\pi_\sigma^{NN}$, which reconstruct the local average matter density and the local amplitude of perturbations respectively. We show that these two fields combined are a statistically optimal probe of $\fnl$ in the squeezed limit, i.e. that adding more fields would not increase the Fisher information. We then define optimal loss functions to learn these fields with neural networks. Overall, this results in an elegant picture, where any small-scale information can be added into the neural network data and used to reduce the noise on the $\pi$ fields and immediately tighten constraints on $\fnl$. In the present work we avoid applying the neural networks to individual halos and instead learn from the local halo population distribution. In future work, we will use a Graph Neural Network to extract even more information from the individual objects, making full use of the power of this formalism.

In the second part of the paper we apply this formalism to the simulations of AbacusSummit and AbacusPNG \cite{10.1093/mnras/stab2482,10.1093/mnras/stab2484,hadzhiyska2024abacuspng}. These simulations have a very high mass resolution as well as a large volume, which allows us to probe the method in the interesting high halo density domain. We demonstrate that our method works on these simulations and recovers unbiased $\fnl$ results. We compare the traditional mass-binned halo bias analysis with our new method and find significant improvements if the neural network observes additional information, in the form of halo concentrations.

\section{The two-field $\pi$-field formalism}

In this section we review the formalism of \cite{Giri_2023} and generalize it to the case where the matter field is not observed.

\subsection{The physics of local non-Gaussianity}

To understand the method intuitively, we briefly recall the well-known physics of local primordial non-Gaussianity. In an $\fnl$ cosmology, the initial conditions of the primordial potential are given by:
\begin{equation}
\Phi(\x) = \Phi_G(\x) + \fnl(\Phi_G(\x)^2 - \langle \Phi_G^2 \rangle)  \label{eq:pbs_fnl}
\end{equation}
A theoretical target, set for observers for example in \cite{Alvarez:2014vva}, says that an $\fnl$ detection of $\fnl \simeq 1$ or larger requires a multifield model of inflation to be explained, while $\fnl \ll 1$ favours single-field inflation. 

\begin{figure}[t!]
\centering
  \includegraphics[width=0.7\textwidth]{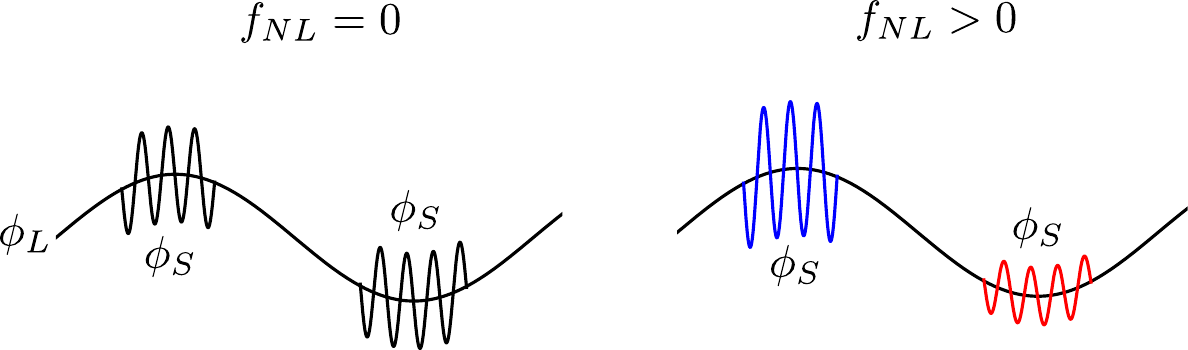}
  \caption{The physics of local non-Gaussianity in the primordial potential $\phi$. Local non-Gaussianity can be visualized as a large scale modulation of small-scale power. Left: We schematically illustrate small scale modes $\phi_S$ in two different regions of the universe, as well as a large-scale mode $\phi_L$ that provides a background for them. With $\fnl=0$ the background mode does not influence the small-scale modes, which would have statistically the same power spectrum at all locations. Right: $\fnl>1$ leads to a coupling of the modes, so that there is more small-scale power where the long mode is large, and less small-scale power where it is small.} 
\label{fig:fnl}
\end{figure}

To analyze the effect of a long-wavelength mode, let us 
decompose the {\em Gaussian} potential as a sum $\Phi_G = \Phi_L + \Phi_S$ of long-wavelength and short-wavelength 
contributions. It can then be shown that the long wave-length modes modulate the power in the small-scale modes, as illustrated in Fig.\ref{fig:fnl}. This can be interpreted as a change in the locally measured amplitude of perturbations $\sigma_8$. The \say{locally observed} value of $\sigma_8$ fluctuates throughout the universe, and is given on large scales by (see \cite{Giri:2023mpg}):
\begin{equation}
\sigma_8^{\rm loc}(\x) = \big( 1 + 2 \fnl \Phi_l(\x) \big) \, \bar\sigma_8 
\label{eq:pbs_sigma8_loc}
\end{equation}
Any probe of local $\sigma_8$ can be used to make an $\fnl$ estimate. In the traditional analysis, called \emph{scale-dependent halo bias} \cite{Dalal:2007cu}, this is done by measuring the local halo density. However, as we shall see, the local halo density is not the optimal probe of local $\sigma_8$. As we have seen, $\fnl$ leads to a large-scale modulation of local $\sigma_8$ (i.e. local primordial power). Intuitively it is thus clear that an optimal probe of $\fnl$ required two ingredients: 
\begin{itemize}
    \item An optimal measurement of $\sigma_8^{loc}$ as a function of position and
    \item an optimal measurement of the long modes $\Phi_L$. 
\end{itemize}
In technical terms, the measurement of the long mode allows for sample variance cancellation, and lowering the noise of the long mode is equivalent to lowering shot-noise. On the very large scales that we are interested in here, bulk flows can be neglected (see App. \ref{sec:shotnoise} for an exploration of bulk flows with respect to shot noise), and we can assume that we are measuring $\sigma_8^{loc}$ and $\phi_L$ at the same comoving position today as in the primordial universe.

\begin{figure}[t!]
\centering
  \includegraphics[width=0.7\textwidth]{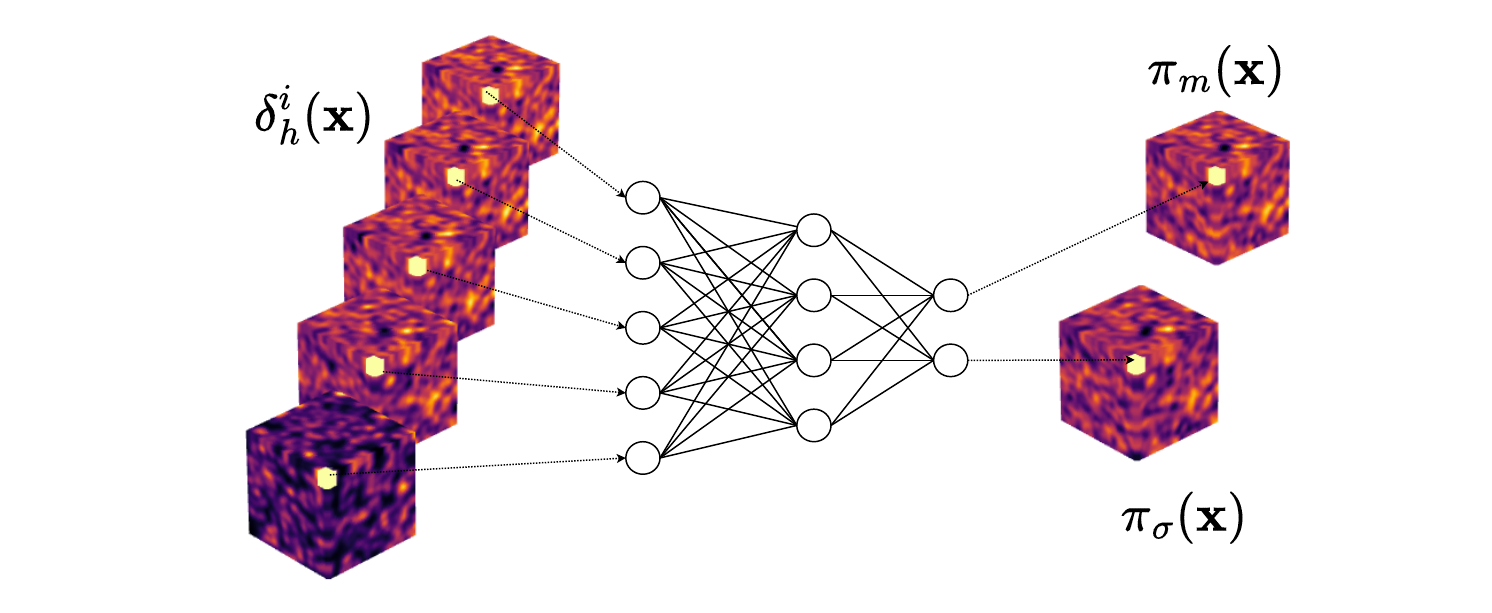}
  \caption{Visualization of the formalism. The input to the NN is derived directly from the halo catalog, mass- and mass-concentration-binned halo density fields in the scope of this paper. The neural net is local (the receptive field is depicted in yellow). The two output fields are built in a way to maximize sensitivity to the $f_{NL}$. One field, $\pi_m$, reconstructs the large-scale linear matter overdensity, the other, $\pi_\sigma$ - the local primordial power spectrum.}

\end{figure}

\subsection{Scale-dependent bias and the $\pi$-field formalism}

A well-known result in cosmology shows that the long wave-length halo field, a biased tracer of the matter field, develops a non-linear bias due to the presence of local primordial non-Gaussianity.

\begin{equation}
    \delta_h(\mathbf{k}_L) = \left( b^G_h+2b^{NG}_h\frac{f_{NL}}{\alpha(k,z)} \right)\delta_m(\mathbf{k}_L)
\end{equation}
where $\alpha(k,z)$ is a coefficient that relates the gravitational potential $\Phi$ to the matter field via Poisson equation $\delta_m(\mathbf{k}) = \alpha(\mathbf{k})\Phi(\mathbf{k})$; $b^G_h$ is a linear halo bias and the second term, $b^{NG}_h\frac{f_{NL}}{\alpha(k,z)}$, a scale-dependent bias due to non-Gaussianity. The expression for $b^{NG}_h$ is given by
\begin{equation}
    b^{NG}_h = \frac{\partial \log \bar{n}}{\partial \log \sigma_8}
\end{equation}
We note here that the value of this bias is defined by the sensitivity of average halo abundance to the $\sigma_8$. The mechanism according to which the biased tracer of the matter field acquires the scale-dependent bias due to non-Gaussianity is a consequence of the equivalence principle. Therefore, any other biased local tracer of the matter field should acquire the scale-dependent bias similarly. 

In \cite{Giri:2023mpg} we showed that any field $\pi$ which is sensitive to the local amplitude of perturbations will have scale dependent bias, i.e. on large scales it can be described as
\begin{equation}
\label{eq:biasmodel}
\pi(\k) = \left(b^G_\pi+2b^{NG}_\pi\frac{f_{NL}}{\alpha(k,z)}\right)\delta_m(\mathbf{k}_L) + \epsilon_\pi(\k)
\end{equation}
where $b^G_\pi$ is its Gaussian bias, $b^{NG}_\pi$ is its non-Gaussian bias and $\epsilon_\pi$ it its Gaussian noise. The $\pi$ field can for example be the halo field (in the traditional method), the local power spectrum, or a neural network derived field. The value of $b^{NG}_{\pi}$ is then given by 
\begin{equation}
\label{eq:bng_pi}
     b^{NG}_{\pi} = \frac{\partial \bar{\pi}}{\partial \log \sigma_8} 
\end{equation}
and can be estimated from simulations (see below for a discussion of uncertainty of $b^{NG}_{\pi}$).

\subsection{Analysis of the $\pi$-field formalism when $\delta_m$ is known}
\label{sec:singlefieldformalism}

In the following we will show hat the neural network derived $\pi$ field is optimal if the loss function is chosen correctly. In this subsection we will first perform the analysis in the case where the matter field $\delta_m$ is known, which simplifies the argument and expressions. In the next section we will drop this assumption. We summarize our arguments here, and defer detailed calculations to the appendices.

\subsubsection{Fisher forecast for $N$ different $\pi$ fields and $\delta_m$}

We first derive a simple formula for the Fisher information contained in a set of N $\pi$-fields (such as the halos in a set of mass bins), together with a known matter distribution $\delta_m$. We assume that the fields obey $\langle\pi_i\pi_j\rangle = b_ib_jP_{mm}+N_{ij}$ and $b_i = b^G_i+\frac{2b^{nG}f_{NL}}{\alpha}$. Then the total covariance takes the following form

\begin{equation}
    \mathbf{C} = \begin{bmatrix}
     P_{mm}(k) & b_iP_{mm}(k) \\
    b_jP_{mm}(k) & b_ib_jP_{mm} + N_{ij} \\
\end{bmatrix}
\end{equation}
The fisher matrix is $F_{ab} = \frac{1}{2}\sum_{k}\mathcal{F}_{ab}(k)$ where $\mathcal{F}_{ab}(k) = \mathrm{Tr}[\mathbf{C}_{,a}\mathbf{C}^{-1}\mathbf{C}_{,b}\mathbf{C}^{-1}]$ and indices $a,b\in\{f_{NL},b^G_{i},N_{ij}\}$. Let's define

\begin{align}
    c_{00} = \sum_{k}\frac{8P_{mm}(k)}{\alpha^2(k)};\ \ \ c_{01} = \sum_{k}\frac{4P_{mm}(k)}{\alpha(k)};\ \ \ c_{11} = \sum_{k}2P_{mm}(k);
\end{align}
We show in App. \ref{app:fisherwithm} that
\begin{equation}
    (\sigma^{2}_{f_{NL}})^{umm} = (F_{f_{NL},f_{NL}})^{-1} = \left[c_{00}b^{nG}_i(N^{-1})_{ij}b^{nG}_j\right]^{-1}
\end{equation}
and 
\begin{equation}
\label{eq:sig_fish_marg}
    (\sigma^{2}_{f_{NL}})^{mar} = (F^{-1})_{f_{NL},f_{NL}} =  \frac{c_{11}}{c_{00}c_{11}-c^2_{01}}\left[b^{nG}_i(N^{-1})_{ij}b^{nG}_j\right]^{-1}
\end{equation}
In this case, where we know large-scale matter field, we thus have the following nice properties:
\begin{itemize}
    \item Noises disentangle from the Fisher matrix - marginalization over noise doesn't affect $\sigma_{f_{NL}}$. 
    \item Gaussian biases don't show up in the expression for $\sigma_{f_{NL}}$.
    \item Marginalized $\sigma_{f_{NL}}$ is equal to unmarginalized one times a factor dependent only on the survey volume.
    \item The expression is exact and holds for any fiducial $f_{NL}$.
\end{itemize}

\subsubsection{Compressing the information to a single field}

Based on the result of the previous section we can define a new single field by
\begin{align}
    \pi' = b_{ng}^T N^{-1} \vec{\pi}
\end{align}
This field has biases
\begin{align}
    b_{g}' = b_{ng}^T N^{-1} b_{g}  \hspace{1cm} b_{ng}' = b_{ng}^T N^{-1} b_{ng}
\end{align}
and noise
\begin{align}
    N' = b_{ng}^T N^{-1} b_{ng} 
\end{align}
We show in App. \ref{app:fisherwithm} that this field contains the same Fisher information as the original N fields. Its Fisher information is 
\begin{equation}
(F^{-1})_{f_{NL}f_{NL}} = \frac{c_{11}}{c_{00}c_{11}-c^2_{01}}\frac{(b'_{ng})^2}{N'}
\end{equation}

\subsubsection{Equivalence of measuring $\sigma_8$ and $\fnl$ and learning the optimal field}

We have shown above that a single optimal $\pi$ field is given by
\begin{align}
    \pi' = b_{ng}^T N^{-1} \vec{\pi}
\end{align}
and that its Fisher information gives an $\fnl$ constraint of
\begin{equation}
\label{eq:fnlfi}
\sigma_{f_{NL}} \propto \frac{(N')^{1/2}}{b'_{ng}}
\end{equation}
We will now show that a field that is constructed to be an optimal probe of $\sigma_8$ is also an optimal probe of $\fnl$ because their Fisher informations are proportional.

Recall that non-Gaussian bias is defined as 
\begin{align}
\label{eq:bng_pi}
b^{NG}_\pi = (\partial\bar\pi/\partial\log\sigma_8)
\end{align}
Thus any field $\pi$ with $b_{ng} \neq 0$ will be sensitive to $\sigma_8$. We can construct a statistic that we will use to constrain $\sigma_8$ as follows
\begin{equation}
\bar\pi = \frac{1}{V} \int_{x} \pi(x)
\end{equation}
where $V$ is the box volume. The statistical error on $\sigma_8$ from this statistic can be calculated to be  
\begin{equation}
\Delta\sigma_8 = 2 \frac{\sigma_8}{V} \left( \frac{N^{1/2}_\pi}{b^{NG}_\pi} \right)
\label{eq:dsigma8}
\end{equation}
which scales in the same way with noise and non-Gaussian bias as the $\fnl$ Fisher information in Eq. \ref{eq:fnlfi}. We have thus found that constraining $\fnl$ is statistically the same as constraining $\sigma_8$. If we can make a field that is maximally good at constraining $\sigma_8$, it will also be maximally good at measuring $\fnl$. 

Therefore, we train a neural network to obtain the best possible measurement of $\sigma_8$. We will use an MSE based loss as is usual in neural network training:
\begin{align}
\label{eq:lossfuncs8}
J &= \left\langle (\bar\pi  - \sigma^{true}_8)^2 \right\rangle_{simulations}  \\
 &=\left\langle\left(\frac{1}{V}\int\pi(\mathbf{x})dV - \sigma^{true}_8\right)^2\right\rangle_{simulations}    
\end{align}
The minimum of this loss function occurs when $\pi$ equals the true $\sigma_8$ in each simulation.

\subsection{Analysis of the two field $\pi$-field formalism when $\delta_m$ is not known}
\label{sec:twofieldformalism}
We now generalize the above analysis to the case where $\delta_m$ is not known. 

\subsubsection{Fisher forecast for $N$ different $\pi$ fields}

In the case when we don't know large-scale matter field, the covariance is as follows:
\begin{equation}
    \mathbf{C} = \begin{bmatrix}
     & &  \\
     & b_ib_jP_{mm} + N_{ij} & \\
     & & \\
\end{bmatrix}
\end{equation}
We can calculate the expression for unmarginalized error as

\begin{align}
 \sigma^{-2}_{f_{NL}} = F_{f_{NL},f_{NL}} = \sum_k \frac{4P^2_{mm}(k)}{\alpha^2(k)}\Bigl[(b^TN^{-1}b^{nG})^2\frac{1-P_{mm}(k)b^TN^{-1}b}{(1+P_{mm}(k)b^TN^{-1}b)^2}
 +\frac{((b^{nG})^TN^{-1}b^{nG})b^TN^{-1}b}{1+P_{mm}(k)b^TN^{-1}b} \Bigr]
\end{align}

\subsubsection{Compressing the information to two fields}

Based on the intuition that we need to optimally measure both the matter field and the local $\sigma_8$ field we construct two fields with properties 
\begin{itemize}
    \item $\pi_m$ with $b_g=1$ and $b_{ng}=0$
    \item $\pi_\sigma$ with $b_g=0$ and $b_{ng}=1$
\end{itemize}
as the linear combination of N $\pi$ fields with minimal possible noise. We derive the weights for these fields in App. \ref{app:fishernomatter}. We also show that the new fields have the same Fisher information as the original N fields. This demonstrates that it is always possible to combine N $\pi$ fields into just two fields that contain the same information, provided that biases and noises are known. An interesting practical application of this result is that one can reduce any set of observed galaxy distributions to two fields, provided that one can estimate their noises and biases. This in turn can help to make an MCMC analysis converge, which can otherwise be diffcult for many fields as we will see below.

\subsubsection{Learning the two optimal fields with a neural network}

We found above that we can construct optimal fields 
\begin{equation}
    \pi^{two} = \begin{pmatrix} \pi_m \\ \pi_\sigma \end{pmatrix}
\end{equation}
with biases
\begin{equation}
    b^{two}_g = \begin{pmatrix} 1 \\ 0 \end{pmatrix} \hspace{1cm} \textrm{and}  \hspace{1cm} b^{two}_{ng} = \begin{pmatrix} 0 \\ 1 \end{pmatrix}
\end{equation}
each with minimal possible noise. 

We now want to find a training method that will result in neural network generated fields with these biases and minimum possible noise (i.e. lower noise than a weighting of halo fields using the weights in Sec. \ref{sec:2fieldweights} would achieve). First note that the first field $\pi_m$ is simply the matter field. This means that an optimal neural network loss is one that reconstructs the matter field as precisely as possible. Using the usual MSE loss we get
\begin{align}
\label{eq:matterloss}
    J = \left\langle \sum_i (\pi_{m,i} - \delta_{m,i}^{true})^2  \right\rangle_{simulations} 
\end{align}
which can be evaluated either in pixel or Fourier space. For neural network convergence it will be useful to low-pass filter the true matter field, so as to reconstruct only the large scales which are needed for the $\fnl$ estimate. We use a top-hat filter with cut-off $k_{max} = 0.05\ \texttt{h/Mpc}$ to low-pass the linear matter field.

The second field we require is $\pi_\sigma$, i.e. a field with $b_g=0$, $b_{ng}=1$ and minimum possible noise. Is this the field that is constructed by minizing the loss in Eq. \eqref{eq:lossfuncs8}? This is not quite the case. While the resulting field $\pi'$ will have minimal $\sigma_8$ error proportional to $\frac{N^{1/2}_\pi}{b^{NG}_\pi}$, it does not have Gaussian bias of zero, i.e. it is not uncorrelated with $\delta_m$ even if $\fnl=0$ (due to non-linear gravitationl evolution). However, if we train a neural network field $\pi'_\sigma$ which minimizes the loss Eq. \eqref{eq:lossfuncs8} we can then define a field with Gaussian bias zero by
\begin{align}
    \pi_\sigma = \pi'_\sigma - b^{\sigma'}_g \pi_m,
\end{align}
i.e. by subtracting the optimal reconstruction of the matter field obtained from loss Eq. \eqref{eq:matterloss}. Thus our optimal $\pi_\sigma$ is a linear combination of $\pi'_\sigma$ and $\pi_m$. Changing the basis of Gaussian fields by such a linear combination does not change the Fisher information.

In summary, we expectd that constructing two fields $\pi_m$ and $\pi'_\sigma$ using loss functions Eq. \eqref{eq:matterloss} and Eq. \eqref{eq:lossfuncs8} leads to optimal fields to constrain $\fnl$. It is not possible to train further fields that would add anything to the Fisher information. It would be interesting to develop a formal proof for the arguments given above, and derive optimality conditions for the loss functions.

\section{Analysis pipeline}

In this section we apply our method to high-resolution halo catalogues from AbacusSummit and AbacusPNG. 

\subsection{Datasets}

To test the sensitivity to $f_{NL}$, we need a large-volume simulation that is large enough in size to probe scale-dependent bias and includes a high halo density.
Fortunately, such simulations exist by now. We will use:
\begin{itemize}
    \item AbacusSummit simulation \cite{10.1093/mnras/stab2482,10.1093/mnras/stab2980,10.1093/mnras/stab2484}. We use CompasSO halo catalogs at redshift $z=0.3$ obtained from simulation of the evolution of $6912^3$ particles in the volume of $(2\ {\rm Gpc/h})^3$. For training we use the simulations that have different $\sigma_8$ parameters, while all other $\Lambda_{CDM}$ parameters were fixed at baseline values of Planck2018. 
    Mainly, we use 25 $\texttt{AbacusSummit\_base\_c000\_ph000-024}$ fiducial simulations with $\sigma_8=0.808$, 6 $\texttt{AbacusSummit\_base\_c004\_ph000-005}$ simulations with $\sigma_8=0.75$ and one $\texttt{AbacusSummit\_base\_c116\_ph000}$ with $\sigma_8=0.866$.
    \item AbacusPNG simulation \cite{hadzhiyska2024abacuspng}. This is a recent extension to the AbacusSummit set that has fiducial Planck18 cosmology but varies the value of local primordial non-Gaussianity $f_{NL}$. This simulation set has lower mass resolution. Unless otherwise specified, $\fnl$ sensitivity below is evaluated with AbacusPNG resolution.
\end{itemize}
AbacusPNG has modest mass resolution
($M_{min,h} = 35.5\times 10^{10} hM_{\odot}$), while AbacusSummit has another factor of five better resolution ($M_{min,h} = 7.4\times 10^{10} hM_{\odot}$).  We generally use AbacusSummit, the larger set with different $\sigma_8$ values, for training the neural network and AbacusPNG to evaluate $\fnl$ sensitivity.

\subsection{Neural network architecture and training procedure}

\subsubsection{Preparing the field data for training}
\label{sec:fielddata}
Our goal in this work is to investigate how various halo features (mass, concentration) can be utilized to maximize the sensitivity to $f_{NL}$. Rather than considering individual halos, for simplicity here we provide the neural network with the local halo population statistics in each voxel. We sample the halo positions on a $128^3$ grid (resulting in 15.625 Mpc/h voxel side length). In each voxel, we provide a number of halo population statistics. In particular we consider the following cases:
\begin{itemize}
    \item Local halo mass distribution in 5 mass bins. Our input array for the neural network is thus of size [128,128,128,5]. By binning the masses broadly we implicitly include some error on their observational precision.
    \item Local halo mass and concentration distribution. Halo concentration $c$ is defined as a ratio of viral to scale radius of the halo profile - $c = r_{vir}/r_s$. As in \cite{rocher2024desi}, we use $r_{98}$ and $r_{25}$ as the proxies for the corresponding radii so that $c_h = r_{98}/r_{25}$. To include the concentrations, we instead bin the halo catalogs in $5\times 4$ mass-concentration bins, as shown in Fig. \ref{fig:nhch} (right). We flatten the mass concentration array, so that the neural network input is of form [128,128,128,20]. 
    \end{itemize}
The dependence of scale-dependent bias on the stellar mass was studied in \cite{Barreira_2020}, and the power of concentrations was demonstrated in \cite{Sullivan:2023qjr, Lazeyras_2023}. In \cite{shao2022robust}, it was shown that halo concentrations are helpful for neural-net-based inference of $\sigma_8$ from dark matter halo catalogs. Concentrations are not directly observable. In the present work we assume concentrations to be known, see \cite{Sullivan:2023qjr} for an exploration of reconstructing concentrations from observable galaxy properties.

Intuitively we expect that in the first case the neural network should be about equally powerful at constraining $\fnl$ as a traditional analysis with the same 5 halo mass bins as five $\pi$ fields. In the second case we expect that the neural network can extract extra information from the concentrations. In principle, one could imagine making $\pi$-fields which bin in both mass and concentration (here 5x4). However, running an MCMC with so many parameters becomes increasingly difficult in practice due to the high shot noises and proliferation of bias parameters. Our machine learning technique therefore can also be valuable if the non-linearity of a neural network is not required and one can use the linear weights of Eq. \eqref{eq:fieldweights1} and Eq. \eqref{eq:fieldweights2}, since these weights require a measurement of many bias and noise parameters from the simulation.

\subsubsection{Neural network architecture and training}
\label{subsec:nn_training}
For the $\sigma^{loc}_8(\mathbf{x})$, we construct the neural network as a simple 3D convolutional neural network (CNN) that has only 1x1x1 convolutional filters so that the resulting receptive field is the same as the resolution - $15.625^3 {\rm Mpc/h}$. In this case it's equivalent to a multilayered perceptron with the input layer dimensionality equal to the number of channels (histogram bins). For matter reconstruction, we found improvement with 3x3x3 convolutions in the first layer. We found that it's sufficient to have very a simple neural net - in all cases we use at most four nonlinear layers with at most 64 neurons per layer (we provide the exact configuration in App. \ref{app:nn_arch}). In fact, non-linearity is not essential in the current setup (but will be required when we upgrade the neural network to a graph neural network, see Sec. \ref{sec:conclusion} for more discussion). 

To isolate the effects of the primordial power spectrum amplitude on the local halo formation, we need a set of simulations with fixed cosmology and variable $\sigma_8$. In most cases we used 6 Abacus simulations with $\sigma_8=0.749$ and 6 with $\sigma_8=0.808$ as our training set, keeping the rest out of the total of 25 simulations with fiducial cosmology to track validation loss. In some cases, we found it beneficial to include one simulation with $\sigma_8=0.866$ in the training set as well (with a weight of 6 to keep it balanced). The neural network $\pi_\sigma^{NN}$ is trained to minimize the squared difference between the true global $\sigma_8$ and volume-averaged prediction as in Eq. \eqref{eq:lossfuncs8}. The neural network $\pi_m^{NN}$ is trained to minimize the loss Eq. \eqref{eq:matterloss}. We use AdamW optimizer with the learning rate $4\times10^{-4}$, weight decay of $1\times10^{-4}$ and a batch size of 3. The overall training time in all cases didn't exceed 40 minutes. In some instances when training $\sigma^{loc}_8$ with only the halo masses, we found that convergence varies from training to training. To mitigate the training convergence effects, we did multiple training runs and chose the model with the lowest noise. 

\subsubsection{Alternative loss function}
Convergence in our bimodal training of $\pi_\sigma^{NN}$ is relatively slow. We conjectured that training would benefit from training data where $\sigma_8$ varies continuously, rather than being bimodal. We thus experimented with an alternative loss function. We can construct a $\sigma_8^{loc}$ target field from the initial conditions as follows. The initial density is first high-pass-filtered, then squared voxel-wise, and finally averaged across some small local volume. 
\begin{equation}
\sigma^{loc}_8(\mathbf{x}) = \left\langle(W_H[\delta^{lin}_m(\mathbf{x})])^2\right\rangle_{\Delta V}
\end{equation}
Here, $W_H$ stands for a high-pass filter with $k>0.5\ {\rm h/Mpc}$ and $\Delta V = (15.625\ {\rm Mpc/h})^3$ This procedure results in the field proportional to $\langle\delta^S_{prim.}\delta^S_{prim.}\rangle$. Then the neural network is trained to minimize the squared voxel-wise difference
\begin{equation}
\label{eq:alternativeloss}
J = \left\langle\left(\pi(\mathbf{x}) - \sigma^{true}(\mathbf{x})\right)^2\right\rangle_{V, batch}    
\end{equation}
This procedure results in significantly faster training (due to the target variability) and equal errors for the $f_{NL}$ prediction.

\subsection{$\fnl$ likelihood pipeline}
\label{sec:likelihood}

We can then apply the trained neural networks on independent test simulations to generate the fields $\pi_\sigma^{NN}$ and $\pi_m^{NN}$. To estimate $f_{NL}$, we use a field-level likelihood as in \cite{Giri_2023}.  The likelihood is given by:
\begin{equation}
    -2 \ln \mathcal{L}(f_{NL},b^G,N|\delta_m,\pi) = \frac{1}{V}\sum_{0<|k|<k_{max}}\left[\mathcal{D}(\mathbf{k})^{\dagger}C(k)^{-1}\mathcal{D}(\mathbf{k}) + V\log \det C(k)\right]
\end{equation}
With covariance given by 
\begin{equation}
    \begin{bmatrix}
        P_{mm}(k) & (b^G+\frac{2b^{nG}f_{NL}}{\alpha(k)})_{j}P_{mm}(k) \\
        (b^G+\frac{2b^{nG}f_{NL}}{\alpha(k)})_iP_{mm}(k) & (b^G+\frac{2b^{nG}f_{NL}}{\alpha(k)})_i(b^G+\frac{2b^{nG}f_{NL}}{\alpha(k)})_jP_{mm}(k) + N_{ij} \\
    \end{bmatrix}
\end{equation}
and data vector 
\begin{equation}
    \mathcal{D} = \begin{bmatrix}
        \delta_m(\mathbf{k}) \\
        \pi_i({\mathbf{k}}) 
    \end{bmatrix}
\end{equation}
We MCMC sample the parameters - $\{f_{NL},\mathbf{b}^G,\mathbf{N}\}$, assuming flat priors. We assume that we can estimate the non-Gaussian bias $b^{nG}$ on simulations with Eq. \eqref{eq:bng_pi} and do not sample it. This is analogous to the traditional scaled-dependent bias analysis where $b^G$ is sampled but $b^{nG}$ is  modelled, for example via the relation $b^{nG} = 2\delta_c(b^G-1)$.

\section{Results}

In this section we first train the neural networks and evaluate their training results. Then we perform a classical mass-binned scale-dependent bias analysis on AbacusPNG to obtain baseline results. Finally we apply the novel neural network method to AbacusPNG.

\subsection{Neural network training}     
\label{sec:res_nn}

\subsubsection{$\sigma_8$ precision after training $\pi_\sigma^{NN}$}
\label{sec:sigma8results}

We train the NN to estimate the mean $\sigma_8$ by minimizing the loss Eq. \eqref{eq:lossfuncs8} on the AbacusSummit simulations. As we have seen, the error on $\sigma_8$ directly translates into the error on $\fnl$. Here illustrate this on 25 fiducial simulations in $(2\ {\rm Gpc/h})^3$ boxes with $\sigma_8=0.808$. These simulations were not in the training data. While all simulations share the same true $\sigma_8$ value, our training procedure ensures that no over-training is happening. Figure \ref{fig:pi_si8_rec_dm}, left, shows the precision of a NN $\pi$-field in two cases where it was trained on the local halo number count as a function of only halo mass $M_h$ (blue histogram) and mass+concentration $M_h,c_h$, as described in more detail in Sec. \ref{sec:fielddata}. As can be seen, halo concentrations help to increase precision by a factor of 3, as compared to the mass-only case. We also obtained equivalent results with the loss function Eq. \eqref{eq:alternativeloss}.

\subsubsection{Matter field reconstruction $\pi_m^{NN}$}
\label{subsec:rec_matter}
\begin{figure}[tbh!]
\centering
\includegraphics[width=0.35\linewidth]{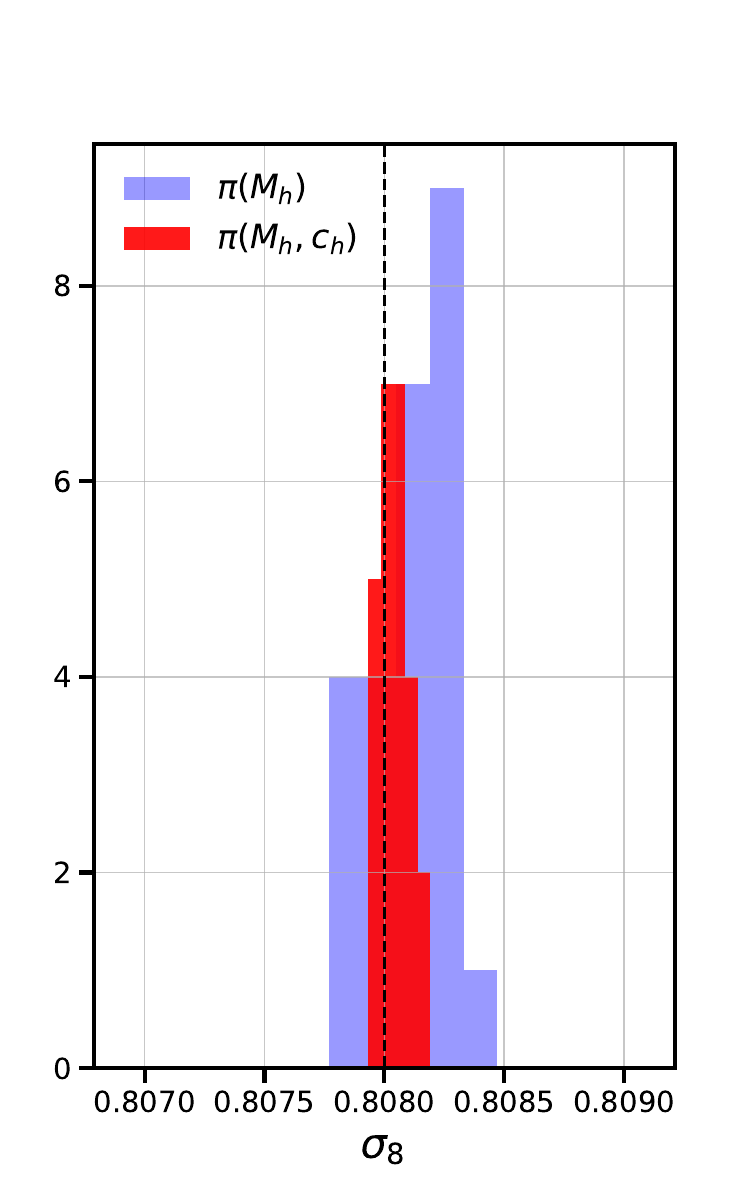}
\includegraphics[width=0.50\linewidth]{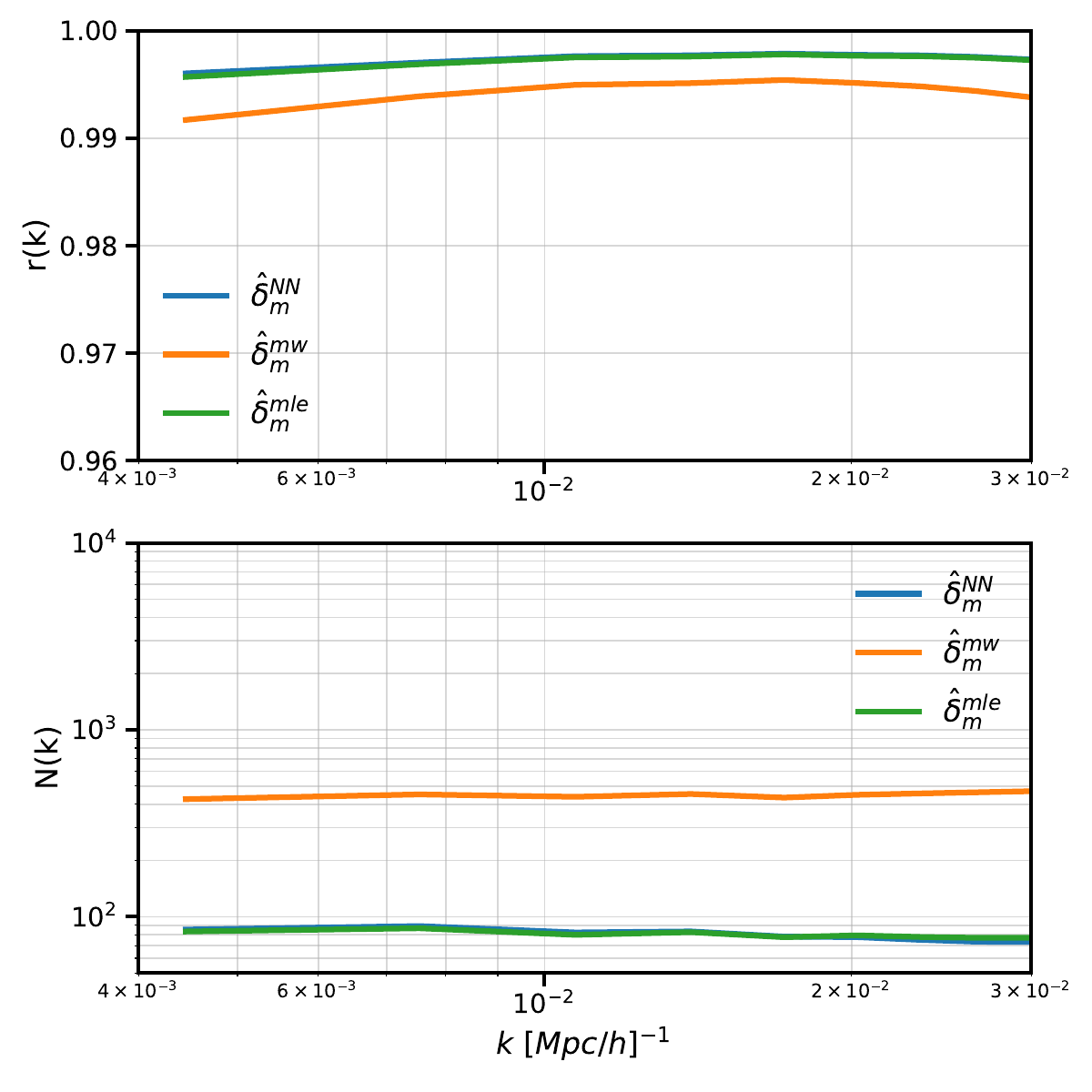}
\caption{Left: The precision in $\sigma_8$ estimation in each of 25 $(2\ {\rm Gpc/h})^3$ volumes of fiducial Abacus simulation with $\sigma_8=0.808$ for a NN trained on local number count as a function of halo mass only (blue) and mass and concentration. The concentrations help to increase precision by a factor of 3. Right: Cross correlation coefficient, $r(k)$, and the power spectrum of the residuals (Noise), $N(k)$ for three estimators of the linear large-scale matter field on the halo catalogs of the fiducial set of AbacusSummit simulation suite. The blue curve corresponds to the NN-based estimator, orange is the mass-weighted halo field and green is linear maximum likelihood estimator (MLE). We find that the neural network performs equivalent to the MLE.}
\label{fig:pi_si8_rec_dm}
\end{figure}

We then train another neural network to reconstruct large-scale linear matter modes. The Abacus simulations do not come with late time matter density, so instead we evaluate the shot noise with respect to the initial conditions, forward projected with 2-LPT. More details about this procedure can be found in App. \ref{sec:shotnoise}. The neural network setup for this task is the same as for local $\sigma_8$ estimation but with minor modifications. The input is the same mass- or mass-concentration- binned halo catalog on a $(128)^3$ grid. The target in this case is low-pass-filtered to $k_{max}=0.05\ {\rm h/Mpc}$ primordial over-density that we evolve to $z=0.3$ with 2-LPT.  The NN is trained to minimize the following loss:
\begin{equation}
    J = \langle (W_L[\hat{\delta}^{NN}_m(\mathbf{x})-\delta^{true}_m(\mathbf{x})])^2\rangle_{batch,V}
\end{equation}
Here $W_L$ stands for low-pass filtering. We train the neural network on 17 fiducial Abacus simulations, keeping the other 8 to track the validation loss. We compare our results to the estimator constructed from binned mass-weighted halo field and to the linear MLE estimator. Figure \ref{fig:pi_si8_rec_dm} (Right) shows the cross-correlation coefficient $r_{xy} = \frac{\langle XY\rangle}{\sqrt{\langle X^2\rangle\langle Y^2\rangle}}$ and the power spectrum of the residuals. We compare the neural network with two traditional methods. In $\delta_m^{mle}$ we use a maximum likelihood reconstruction based on the 5 mass-binned halo fields with weights derived in App \ref{app:mlematter}. In $\delta_m^{MW}$ we mass-weighted each halo by the central mass of its mass bin.

We do not find any significant difference in both NN estimators, trained on either halo mass- or mass-concentration fields. They perform equally well as linear MLE estimator, however, both do better than mass-weighted halo field estimator. In particular, unlike the $\sigma_8$ case, we find that halo concentrations cannot improve the mass reconstruction. Nevertheless, an advantage of using the NN-based estimator, as compared to the MLE estimator is that it doesn't require the estimation of parameters such as biases and noises of the fields. Moreover, the NN-based estimator is more flexible and can incorporate additional information that is either impossible or less convenient to make use of in the traditional approach, in particular different halo properties (positions, velocities, etc.) on an individual basis (see Sec. \ref{sec:conclusion} for more discussion of this point).

\subsection{Conventional scale-dependent halo bias analysis of AbacusPNG}
\label{sec:results_classical}
We first develop a conventional scale-dependent bias pipeline to establish baseline results to compare with the neural network. AbacusPNG has a large halo density, and it is well-known that a single mass-bin analysis is sub-optimal in such a case, in particular since it does not allow for sample-variance cancellation. The neural network will have access to halo masses, so for a fair comparison we need to include mass binning. 

Assume we have split the halo field into $N$ mass bins $\delta_{h,i}$. The auto and cross power spectra on large scales are then given by
\begin{align}
  P_{hh,ij}(k) &= \left( b_{h,i} +\fnl \frac{\beta_i}{\alpha(k)} \right) \left( b_{h,j} +\fnl \frac{\beta_j}{\alpha(k,z)} \right) P_{mm}(k) 
\end{align}
and the covariance is given by 
\begin{align}
    C_{hh,ij}(k) = P_{hh,ij}(k, z) + N_{hh,ij}(k)
\end{align}
where the shot noise $N_{hh,ij}$ is assumed to be flat but may be correlated between bins. We suppress the redshift dependence of the quantities and evaluate the result at redshift $z=0.3$ below.

\begin{figure}[tbh!]
\centering
\includegraphics[width=0.8\linewidth]{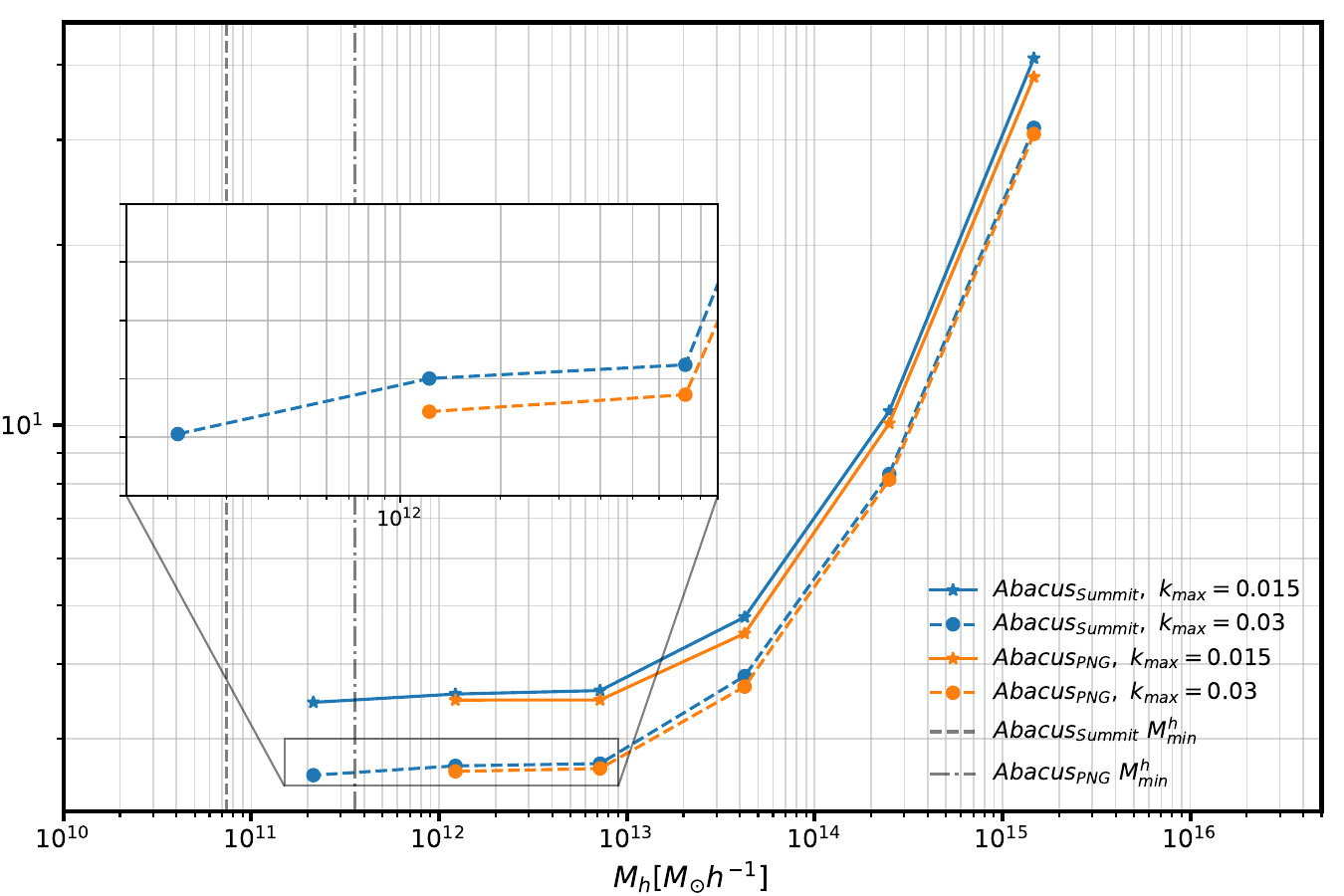}
\caption{$\sigma_{f_{NL}}$ as a function of halo mass cutoff in a traditional mass-binned halo MCMC analysis (Sec. \ref{sec:results_classical}) of halo catalogs from different AbacusSummit (blue) and AbacusPNG (orange) simulations. Solid and dashed lines indicate different Fourier mode cut-offs. The vertical lines indicate the lowest mass in AbacusSummit and AbacusPNG. This analysis assumes that the mass field is known.}
\label{fig:sigma_fnl_fisher_mcmc}
\end{figure}

\begin{figure}[ht]
\centering
\includegraphics[width=0.45\linewidth]{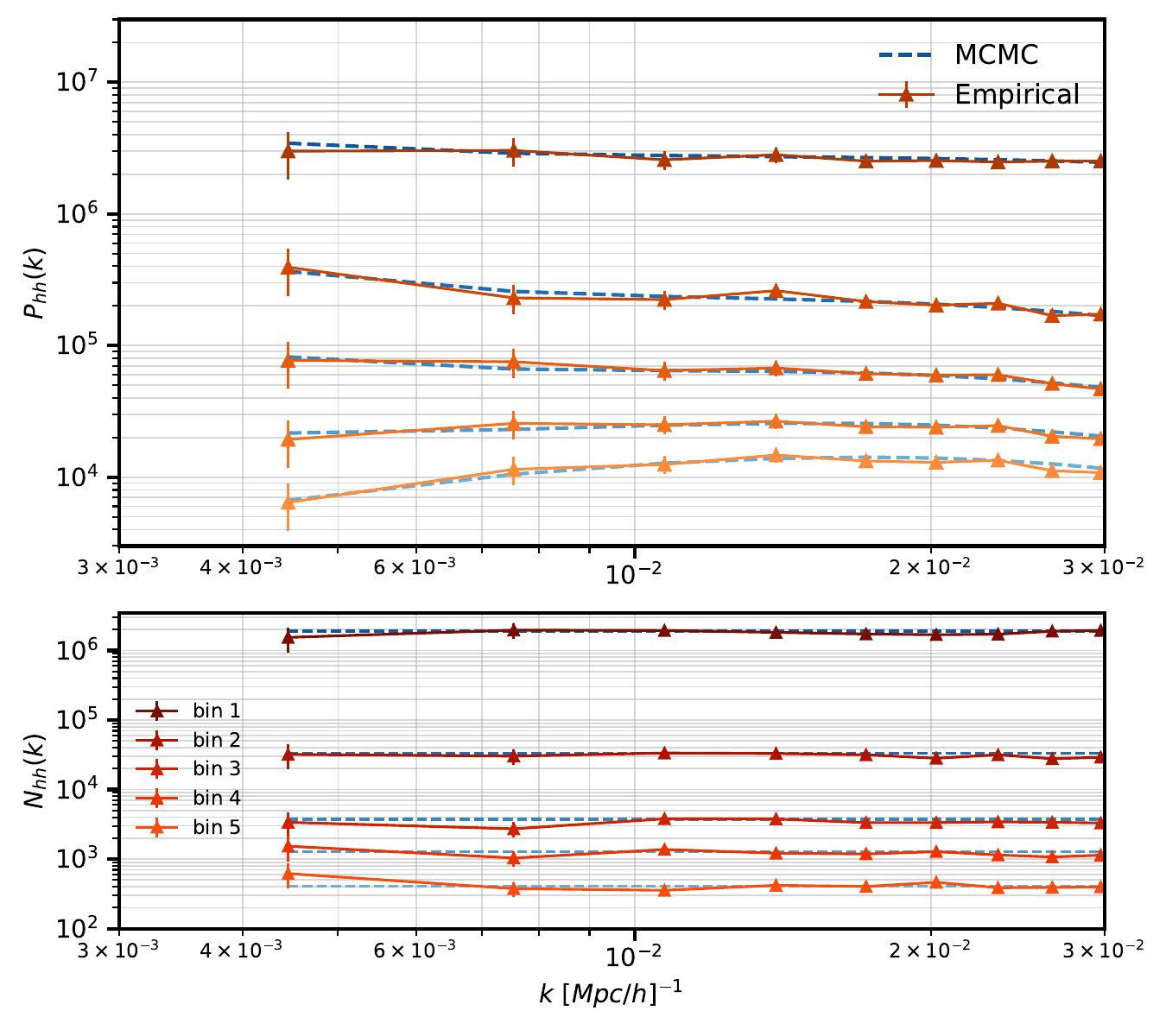}
\includegraphics[width=0.45\linewidth]{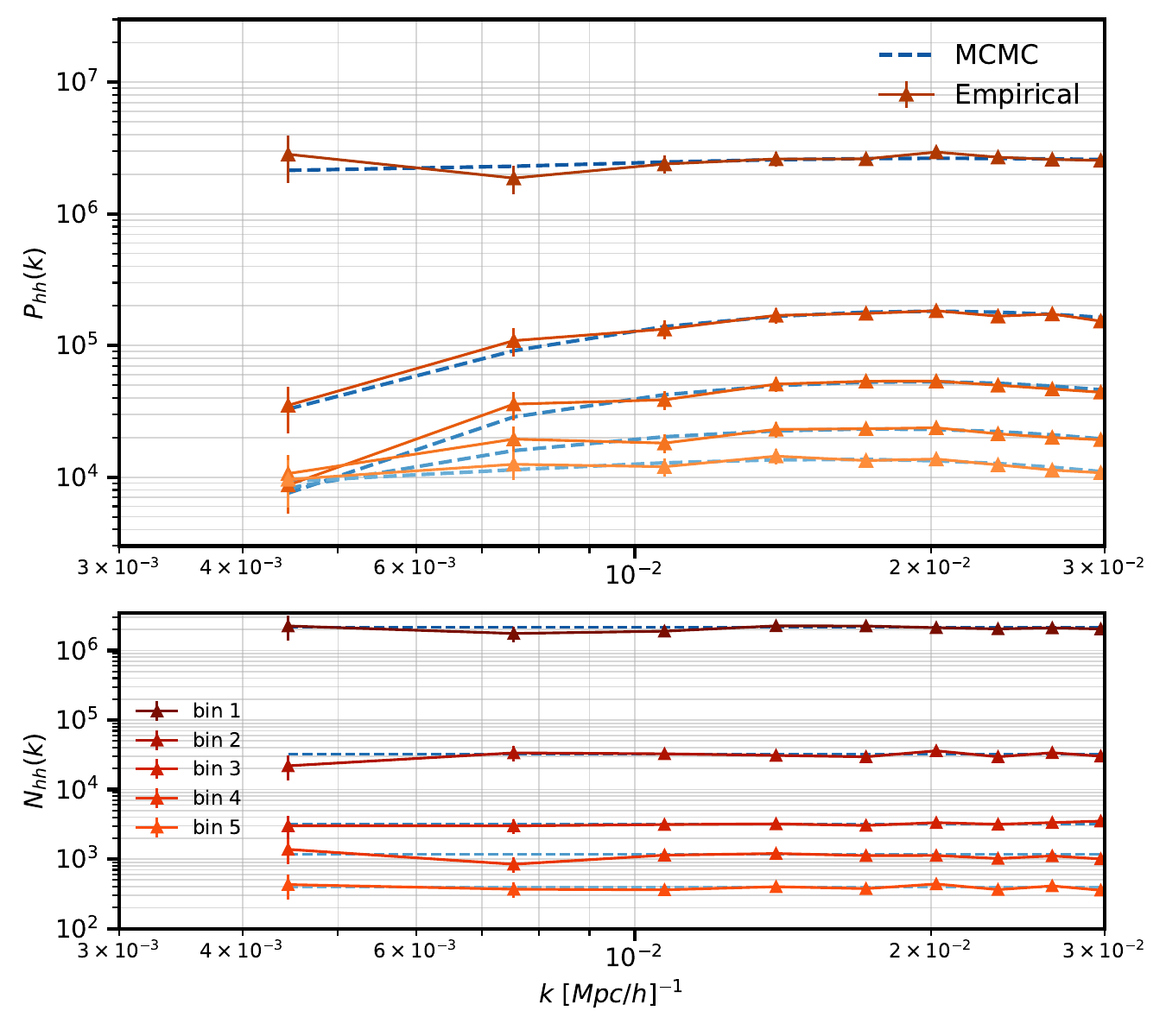}
\caption{Comparison of the large-scale bias model Eq. \eqref{eq:biasmodel} with simulation data for a five mass bin analysis with traditional scale-dependent halo bias (Sec. \ref{sec:results_classical}) on AbausPNG simulations with $f^{true}_{NL}$ = 100 (left) and $f^{true}_{NL}$ = -100 (right). The model fits the simulation data excellently.}
\label{fig:halo_mcmc_fit}
\end{figure}

For the MCMC analysis, we assume the likelihood described in the Sec. \ref{sec:likelihood}. As our data vector we have one large-scale matter field $\delta_m(\mathbf{k})$ and $N=\{1..5\}$ mass-binned halo overdensity fields $\delta_{h,i}(\mathbf{k})$. We limit our analysis to $\mathbf{k}$-modes that are between $k_{min} = \frac{2\pi}{2000}\ {\rm h/Mpc}$ and $k_{max} = \{0.015, 0.03\} \ {\rm h/Mpc}$. The Figure \ref{fig:sigma_fnl_fisher_mcmc} shows the dependence of $\sigma_{f_{NL}}$ from the lowest accessible halo mass on AbacusSummit and AbacusPNG simulation sets. For all different $f_{NL}$ values of AbacusPNG, we find the same $\sigma_{f_{NL}}$ (in agreement with Eq. \ref{eq:sig_fish_marg}), so there's only one curve that corresponds to an average over all the simulations. The AbacusPNG simulation set has higher mass resolution - because of that the blue curves have one more additional point. It turns out to be just enough to probe the beginning of the sample-variance cancellation regime (see e.g. \cite{Ferraro:2014jba}), where $\fnl$ constraints start to improve again with lower halo mass. In Figure \ref{fig:halo_mcmc_fit} we show that the large-scale bias model of Eq. \ref{eq:biasmodel} with  MCMC-derived parameters fits the simulation data on two AbacusPNG simulations with $f_{NL} = {100,-100}$, left and right, correspondingly. For a comparison with predictions from the halo model see App. \ref{app:halomodel}.

\subsection{Neural network $\pi$-field scale-dependent halos bias on AbacusPNG}

We now show the numerical results with our machine learning formalism. We first consider the experimentally unrealistic but computationally simpler case of a known matter field, corresponding to the single field formalism of Sec. \ref{sec:singlefieldformalism}, for two different choices of observed halo properties. We then analyze the realistic case where only halos are observed, corresponding to the two field formalism of Sec. \ref{sec:twofieldformalism}.
 
\subsubsection{Estimation of $f_{NL}$ with halo mass information with matter field known}
\label{sec:results_massknown_massonly}

\begin{figure}[ht]
\centering
\includegraphics[width=0.45\linewidth]{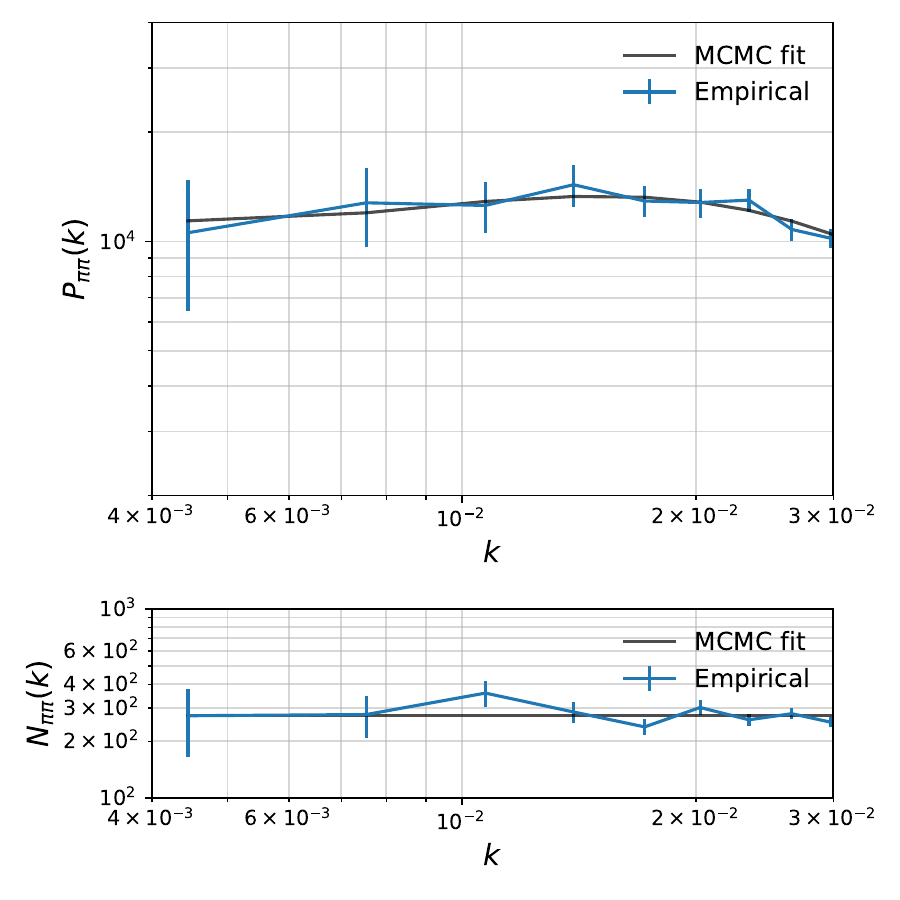}
\includegraphics[width=0.45\linewidth]{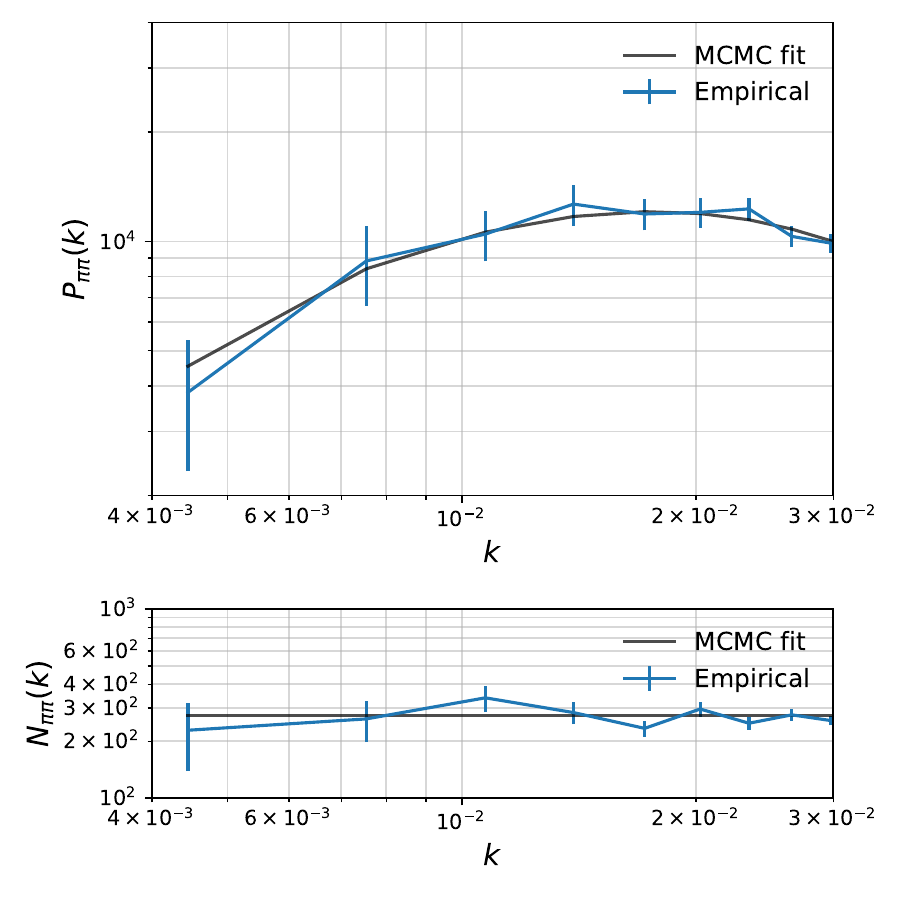}
\caption{Empirical and theory power spectra of $\pi(M)$-fields evaluated on one AbausPNG simulations with $f^{true}_{NL}$ = 30 (left) and $f^{true}_{NL}$ = -30 (right), for the analysis in Sec. \ref{sec:results_massknown_massonly}. Values of $b^G$, $N_{\pi\pi}$ and $f_{NL}$ are obtained from MCMC analysis.}
\label{fig:model_fit_mo}
\end{figure}

First we present the results in the case where we assume known $\delta^L_m$ and the neural network for $\pi$-field was trained only on mass-binned halo data (i.e. without concentration). We consider all five sets of AbacusPNG simulations to run our MCMC analysis on with $f_{NL} = \{-100,-30,0,30,100\}$, 2 simulations for each value of $f_{NL}$, and report the average of 2 runs. For the $k_{max}$, we chose two values - $k_{max} = \{0.015, 0.03\}\ {\rm h/Mpc}$. In all five cases we can recover unbiased $f_{NL}$ with the standard deviation of $\sim 4$ and $\sim 3.3$ for the lower and higher Fourier mode cut-off correspondingly. The results are summarized in the Table \ref{tab:fnl_precision_1f} (left). Figure \ref{fig:model_fit_mo} shows examples of the MCMC model fit, illustrating that the large-scale bias model Eq. \eqref{eq:biasmodel} fits the data.

\begin{table}[tbh!]
\label{tab:fnl_mcmc_1f}
\centering
\begin{minipage}{0.45\linewidth}
\centering
\begin{tabular}{|c|c|c|c|c|}
sim, $f_{NL}$&\multicolumn{2}{c|}{$\hat{f}_{NL}$} & \multicolumn{2}{c|}{$\sigma_{\hat{f}_{NL}}$} \\
\hline
PNG -100   & -95.6 & -97.8 & 4.1 & 3.3  \\
PNG -30    & -25.4 & -27.6 & 4.  & 3.3    \\
PNG 0      & 5.1  & 2.8     & 4.1 & 3.1    \\
PNG +30    & 33.5 & 31.6   & 4.2   & 3.3    \\
PNG +100   & 100.4 &  97.7 & 4.3  & 3.2   \\
\hline
$k_{max}$ & 0.015 & 0.03 & 0.015 & 0.03 \\
\end{tabular}
\end{minipage}
\hfill
\begin{minipage}{0.45\linewidth}
\centering
\begin{tabular}{|c|c|c|c|c|}
sim, $f_{NL}$&\multicolumn{2}{c|}{$\hat{f}_{NL}$} & \multicolumn{2}{c|}{$\sigma_{\hat{f}_{NL}}$} \\
\hline
PNG -100 & -97.7 & -97.7 & 1.2 & 1. \\   
PNG -30  & -29.5 & -29.2 & 1.2 & 0.9 \\
PNG 0    & 0.1   & 0.     & 1.2 & 0.9 \\
PNG +30  & 29.   & 28.6   & 1.2 & 0.9   \\
PNG +100 & 96.3  & 95.9   & 1.2 &  1. \\
\hline
$k_{max}$ & 0.015 & 0.03 & 0.015 & 0.03 \\
\end{tabular}
\end{minipage}
\caption{Results of the MCMC estimation of $f_{NL}$ on five sets of AbacusPNG simulations with true values of $f_{NL} = \{-30,-100,0,30,100\}$ with known large-scale matter field $\delta^L_m$ and NN $\pi$-field trained on mass- (left) and mass-concentration (right) binned halo catalogs. The results are averaged over two simulations in each PNG subset}
\label{tab:fnl_precision_1f}
\end{table}

We find about the same $\fnl$ sensitivity to the traditional analysis using 5 halo mass-binned fields. Our results confirm the expectation that the neural network should recover the same $\fnl$ information as in the original halo-mass binned analysis, because both analyses use exactly the same halo information. The small difference in the sensitivity compared to the traditional result is likely because the neural network is not perfectly converged, due to the limitation of our training data which provides only three different $\sigma_8$ values and only a few simulations. This section is thus a successful consistency check of our method. 

\subsubsection{Estimation of $f_{NL}$ with halo mass and concentration information with matter field known}
\label{sec:results_massknown_massconc}

In Section \ref{sec:res_nn}, we saw how adding the information about halo concentrations helps to train a better estimator of $\sigma_8$. Eqs. \eqref{eq:dsigma8} and \eqref{eq:fnlfi} show that the improvement on the precision of $\sigma_8$ estimation should translate directly to an equivalent improvement on the estimation of $f_{NL}$. Now we test this expectation with MCMC $f_{NL}$ estimation. The setup considered here is analogous to the previous subsection, but now using the local mass-concentration histogram as explained in Sec. \ref{sec:fielddata}.

\begin{figure}[tbh!]
\centering
\includegraphics[width=0.45\linewidth]{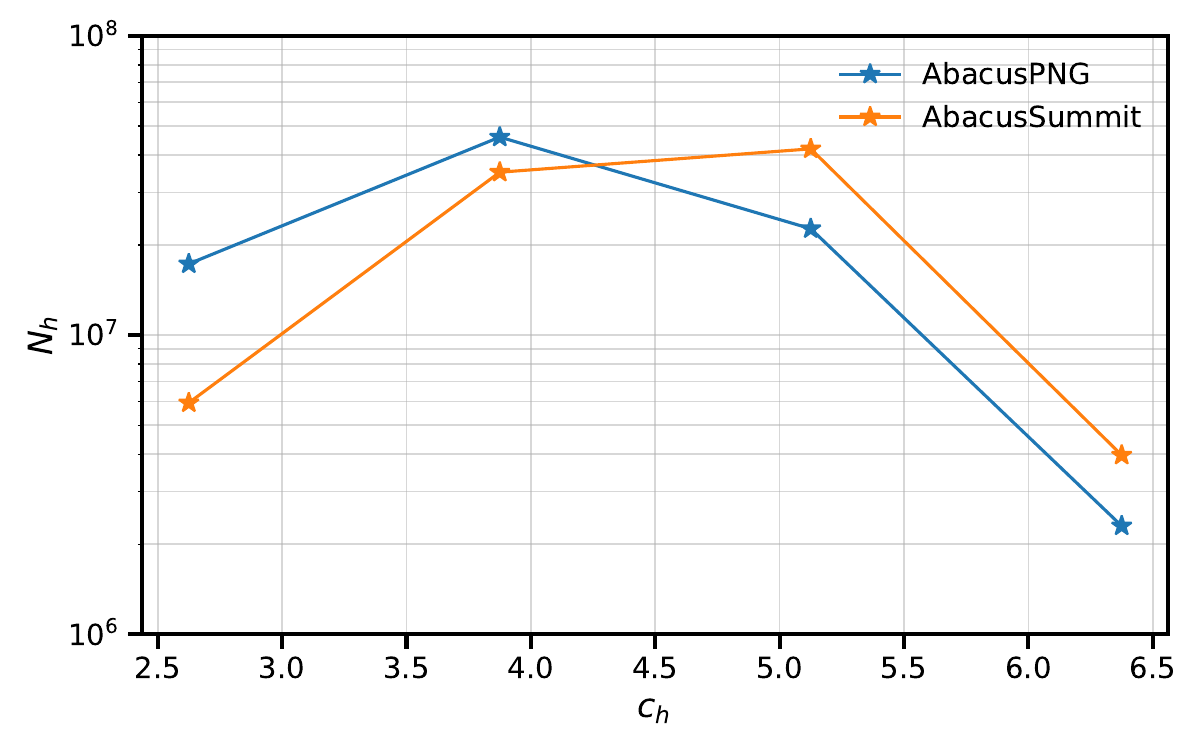}
\includegraphics[width=0.45\linewidth]{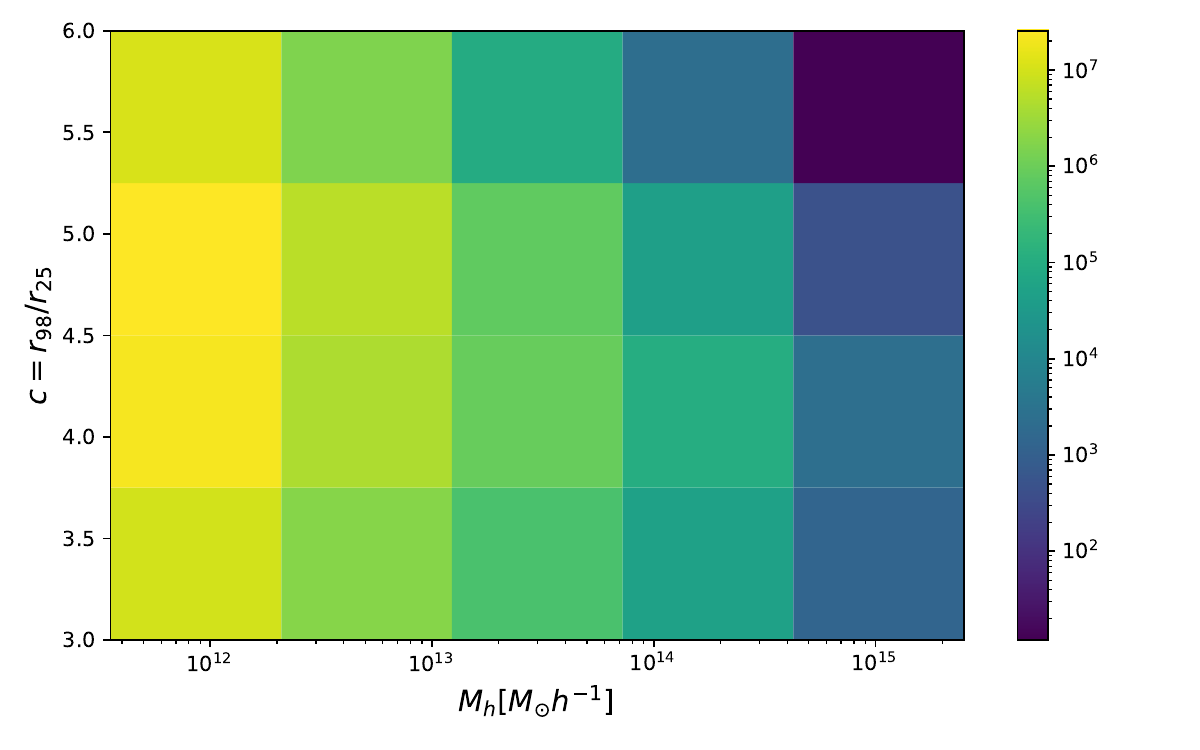}
\caption{Left: Number count of halos in 5 concentration bins in two different simulations, AbacusPNG and AbacusSummit, at the same fiducial cosmology. The same lower halo mass cutoff $\log_{10} M_{min,h}[M_{\odot}h] = 11.5$ has been applied to both simulations. Right: An example of a mass-concentration histogram constructed from the full AbacusSummit simulation. The position-dependent version of this histogram is used as an input to the neural network.}
\label{fig:nhch}
\end{figure}

We initially found somewhat biased $\fnl$ results, which we traced to the following issue. The mass resolution of AbacusPNG simulation set is $\sim 5$ times higher than the one of AbacusSummit, which we used for training. Due to this fact, the distribution of halo number count as a function of concentration is different, as can be seen on the Figure \ref{fig:nhch} (left). To account for this difference between simulation sets, we applied a channel-wise multiplicative input correction when evaluating the NN on AbacusPNG, so that the input data has a mean which is consistent with the training. Ideally, one would like the training simulations (with different $\sigma_8$) and test simulations (with different $\fnl$) to have exactly the same properties, but such a matching data set at high mass resolution is not currently available.

\begin{figure}[ht]
\centering
\includegraphics[width=0.45\linewidth]{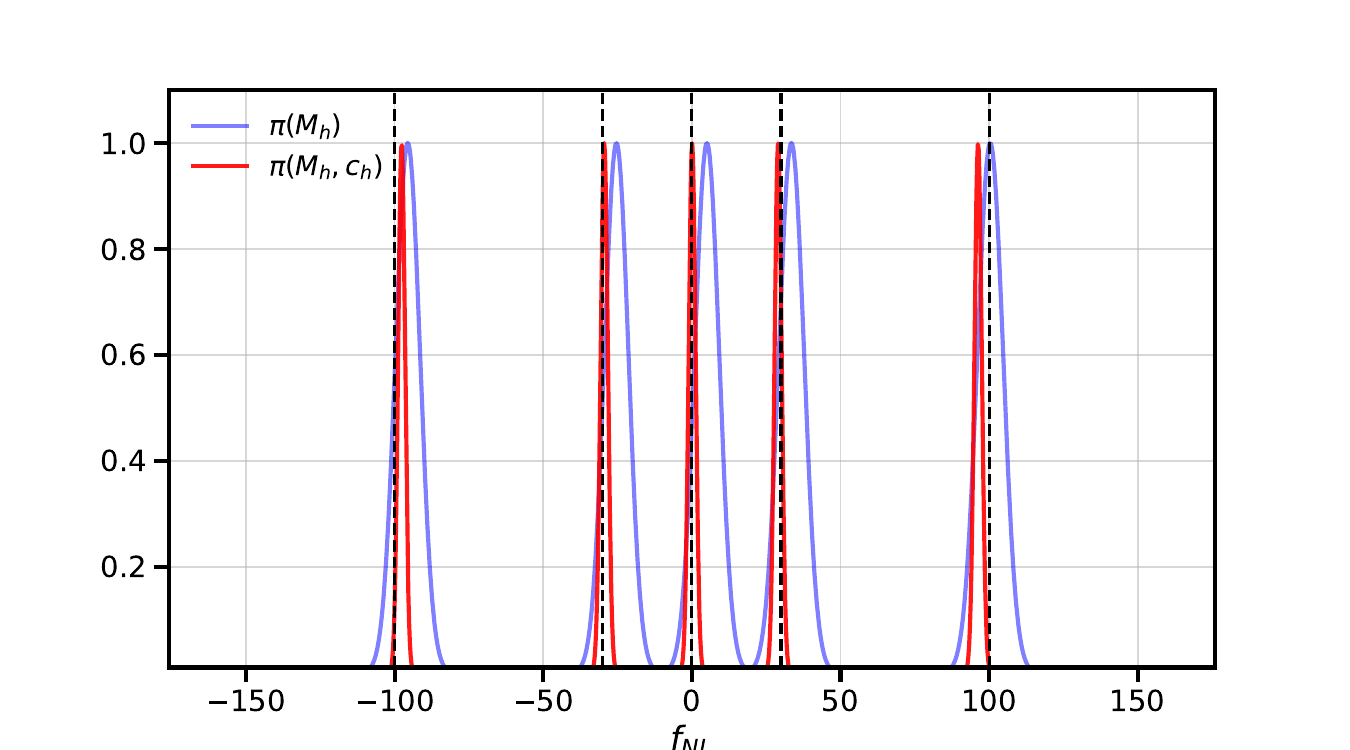}
\includegraphics[width=0.45\linewidth]{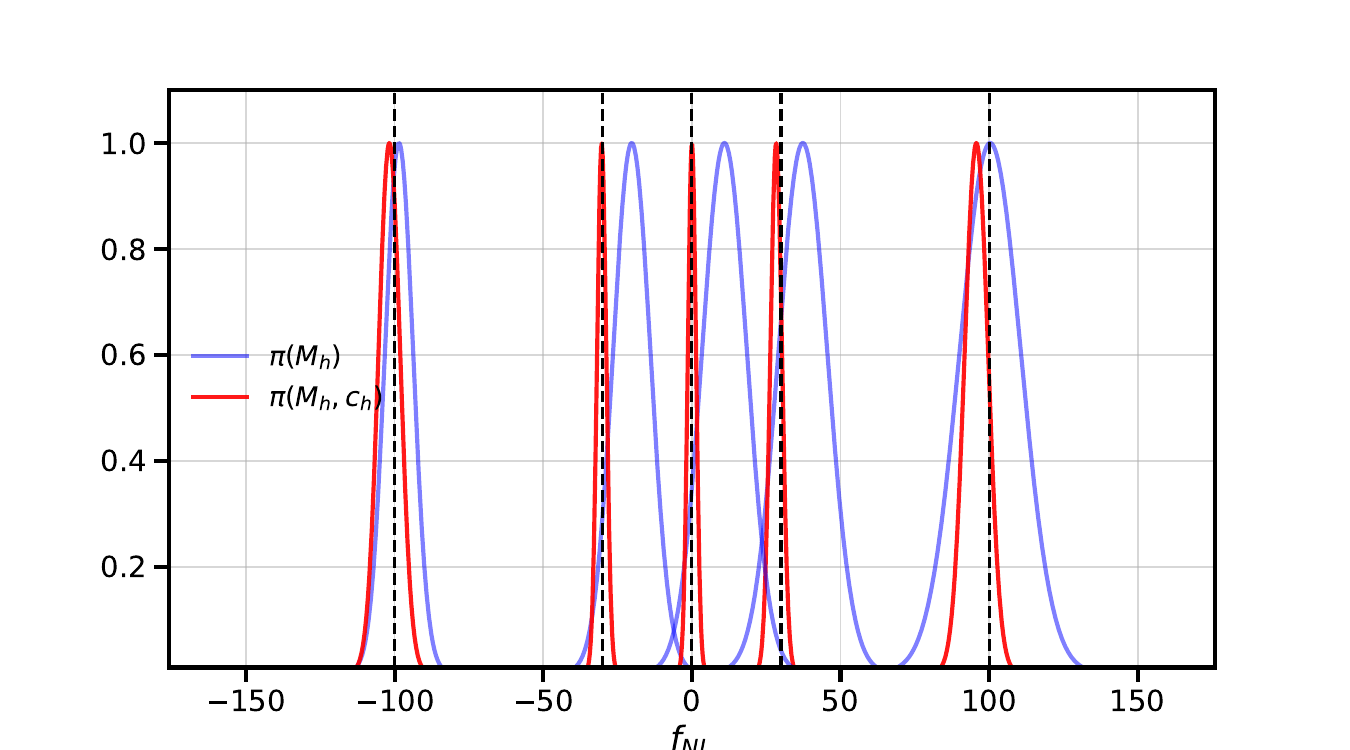}
\caption{Posteriors for $\fnl$ estimation on AbacusPNG with 5 mass bins (blue) and with 20 mass-concentration bins (red). The left plot assumes known large-scale linear matter field (Sec. \ref{sec:results_massknown_massonly} and \ref{sec:results_massknown_massconc}); the right plot shows results with reconstruction of $\delta_m$ (Sec. \ref{sec:results_massunknown}). We obtain unbiases results in all cases.}
\label{fig:sig_fnl_mo_mc_0015}
\end{figure}

\begin{figure}[ht]
\centering
\includegraphics[width=0.45\linewidth]{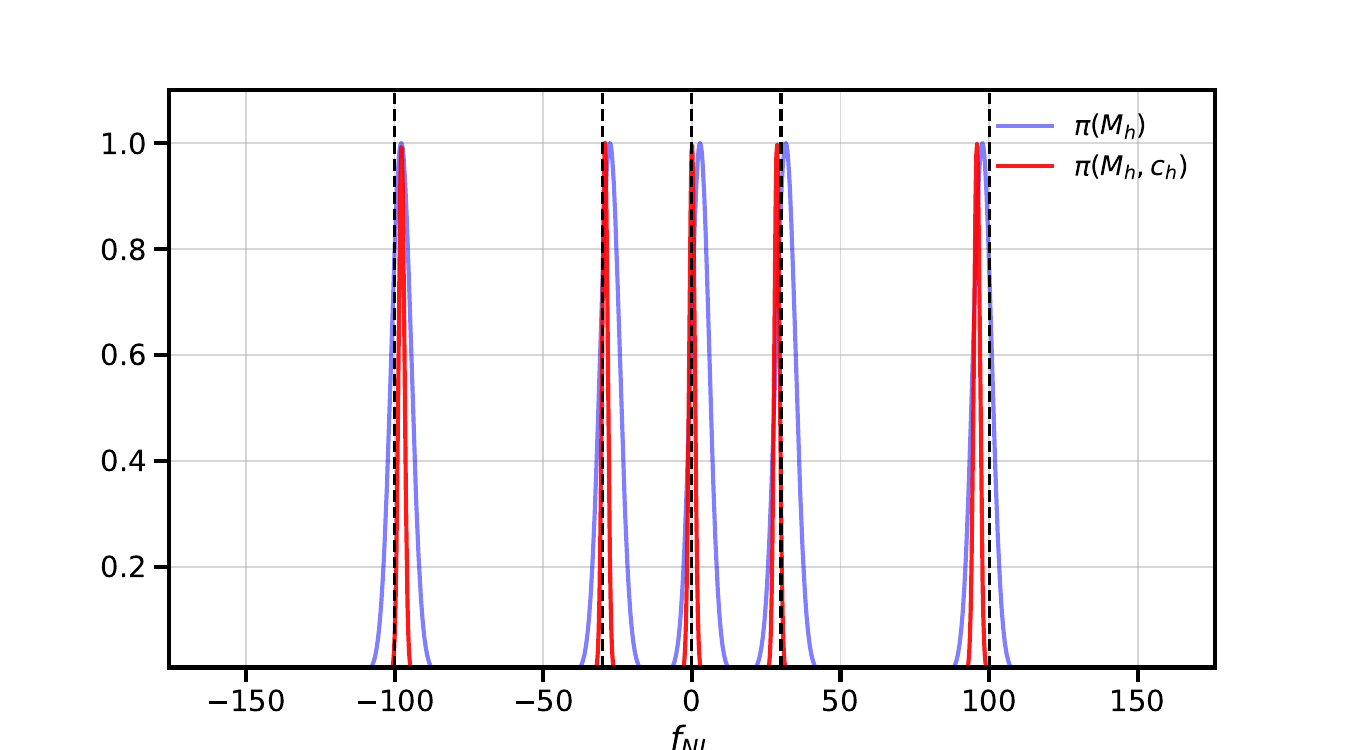}
\includegraphics[width=0.45\linewidth]{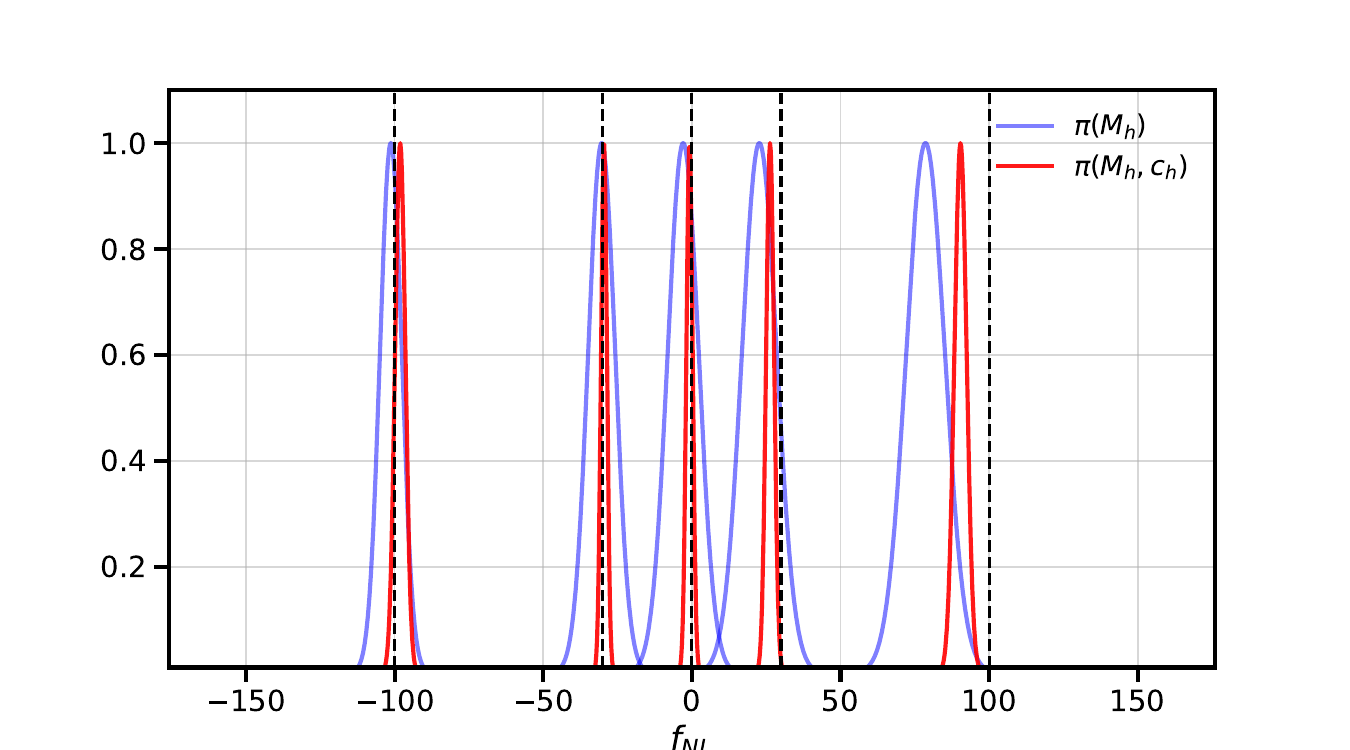}
\caption{Same as Fig \ref{fig:sig_fnl_mo_mc_0015}, but with $k_{max}=0.03\ {\rm h/Mpc}$. In this case there is a slight bias in the $\fnl=100$ data point, indicating that we may be somewhat over the allowed $k_{max}$ range.}
\label{fig:sig_fnl_mo_mc_0030}
\end{figure}

After this de-biasing, we found the results of the MCMC analysis of $f_{NL}$ with the use of $\pi$-field trained on halo mass-concentration function show in the Table \ref{tab:fnl_mcmc_2field} (right). We find that with concentrations our results improve by a factor of $3.5$ in $\fnl$ sensitivity, in agreement with the improvement of 3 in $\sigma_8$ sensitivity found in Sec. \ref{sec:sigma8results}, when comparing to the mass only case. Figures \ref{fig:sig_fnl_mo_mc_0015}, left and \ref{fig:sig_fnl_mo_mc_0030} (left) visualize the $f_{NL}$ sensitivity one can get with $\pi(M_h)$ and $\pi(M_h,c_h)$ for two different cut-off scales $k_{max} = \{0.015, 0.03\}\ {\rm h/Mpc}$. 

\subsubsection{Estimation of $f_{NL}$ with two $\pi$ fields with matter field unknown}
\label{sec:results_massunknown}
\begin{figure}[tbh!]
\centering
\includegraphics[width=0.8\linewidth]{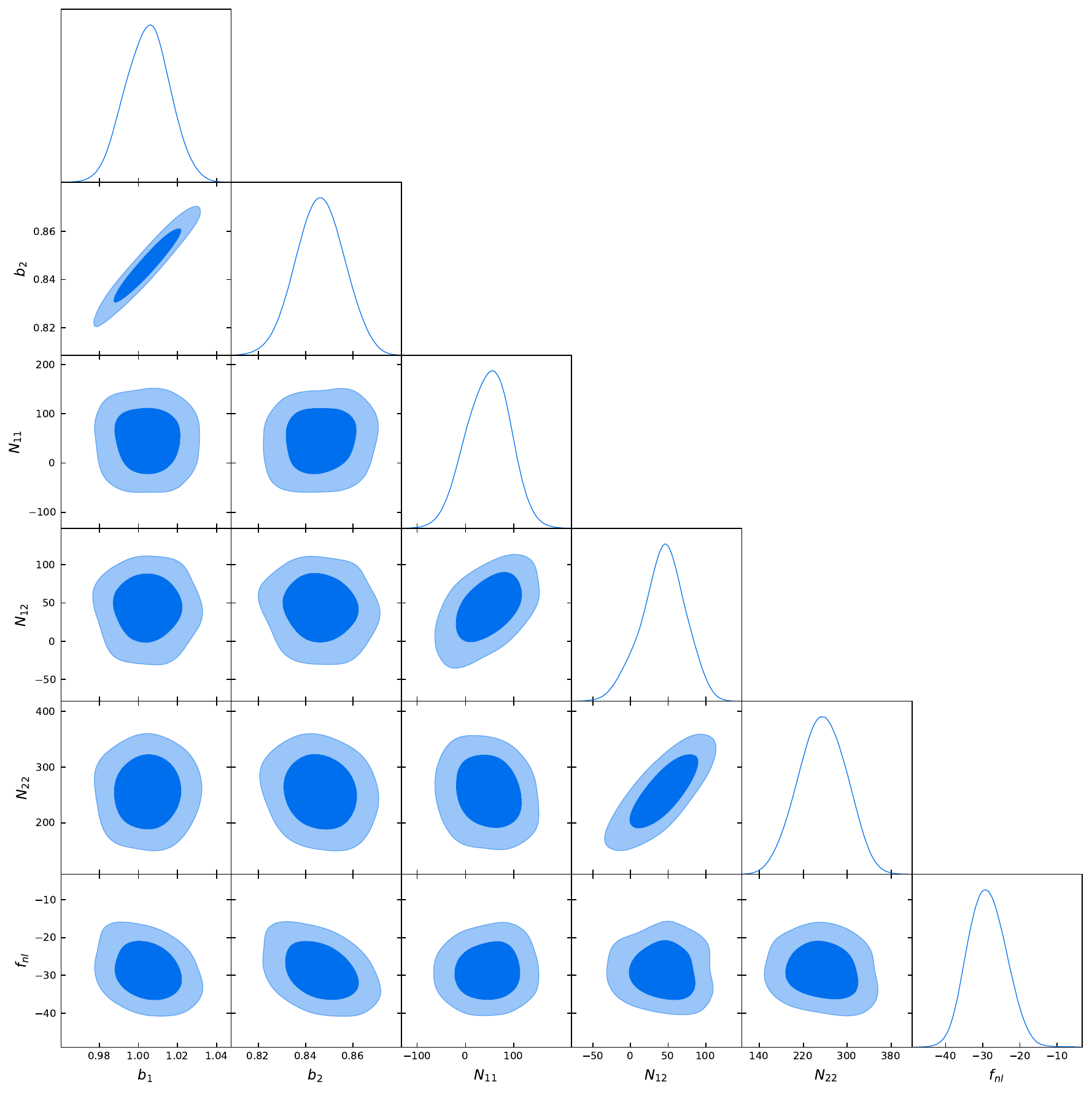}
\caption{MCMC corner plot of $f_{NL}$ analysis of \texttt{AbacusPNG\_base\_c301\_ph000} simulation ($f_{NL}=-30$) with two fields $\pi_m$ and $\pi_\sigma$ trained on local halo mass function $n_h(M)$, as described in Sec. \ref{sec:results_massunknown}.}
\label{fig:corner_2pi_mo}
\end{figure}

\begin{figure}[tbh!]
\centering
\includegraphics[width=0.8\linewidth]{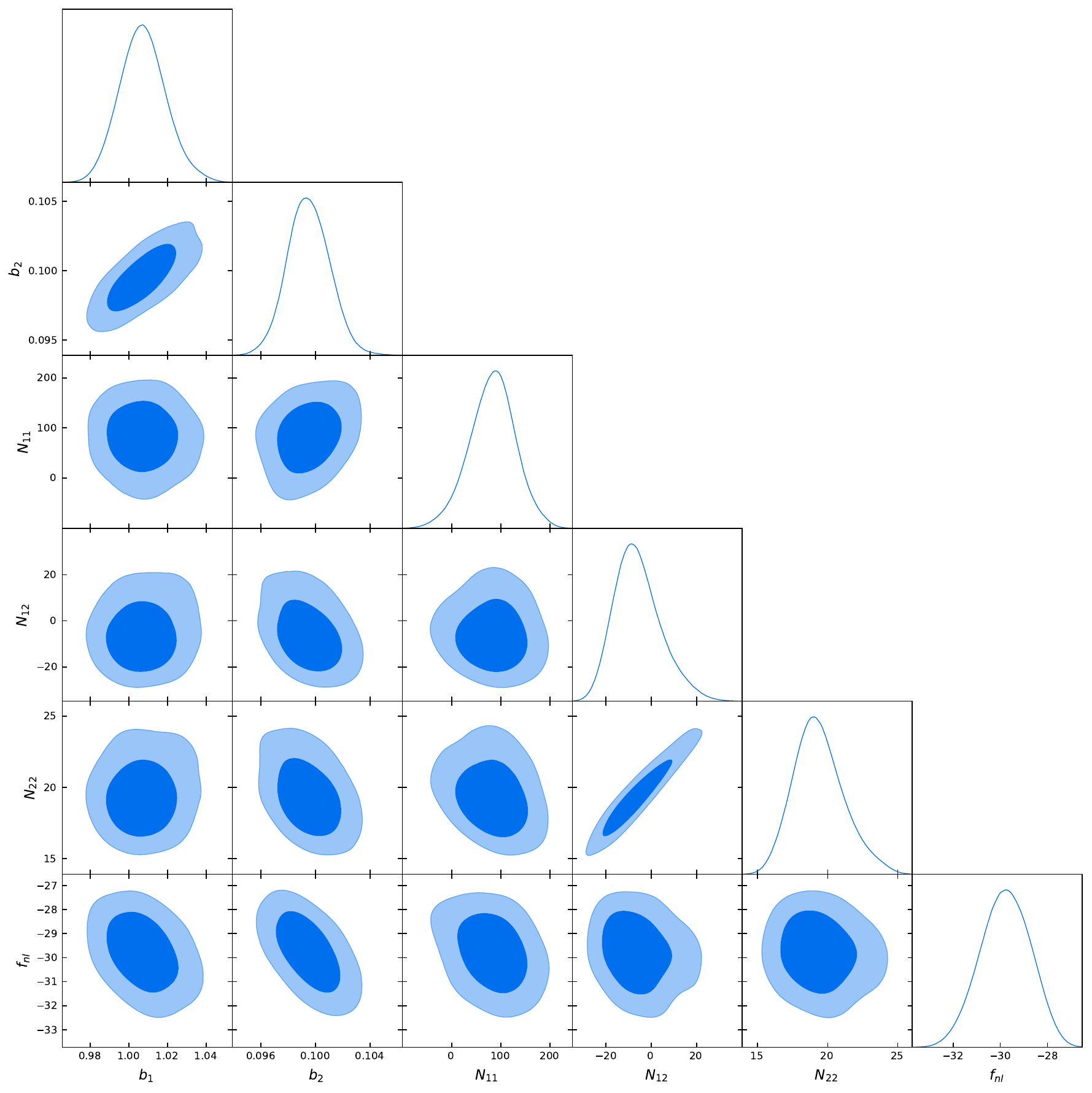}
\caption{Same as Fig. \ref{fig:corner_2pi_mo}, but with $\pi$ fields trained on halo mass-concentration function $n(M,c)$, as described in Sec. \ref{sec:results_massunknown}.}
\label{fig:corner_2pi_mc}
\end{figure}

In the two previous subsections, we assumed that the large-scale linear matter field is known. Here we discuss a more realistic situation, when the matter field is not known but instead reconstructed with the help of the NN, as discussed in Subsec. \ref{subsec:rec_matter}. We consider now two NN fields: $\pi_m$ and $\pi_{\sigma}$ with $\langle\pi_i\pi_j\rangle = (b^{G}_i + 2\frac{b^{nG}_i f_{NL}}{\alpha})(b^{G}_j + 2\frac{b^{nG}_j f_{NL}}{\alpha})P_{mm} + N_{ij}$. We have to include the non-Gaussian bias $b^{nG}_m$ for the reconstructed matter field in the model because in the case of $f_{NL}\neq0$, we reconstruct a field with a scale-dependent bias. We estimate the matter non-Gaussian bias on simulations by evaluating eq. \ref{eq:bng_pi} numerically. As for the Gaussian bias of the reconstructed matter field, the training procedure sets it to the value of 1 by design. However, we still keep it free in the MCMC analysis. Overall, we sample 6 parameters in total - $\{b^{G}_{m},b^{G}_{\sigma},N_{11},N_{12},N_{22},f_{NL}\}$. As in the previous discussions, we compare two models: the one that was trained only on halo mass distributions and the model that has also access to information about halo concentrations. Results of the MCMC analysis are listed in Table \ref{tab:fnl_mcmc_2field} and displayed in Fig \ref{fig:sig_fnl_mo_mc_0015} and \ref{fig:sig_fnl_mo_mc_0030}, right. Examples of the MCMC corner plots are in Fig. \ref{fig:corner_2pi_mo} and Fig. \ref{fig:corner_2pi_mc}. We find very tight constraints on $\fnl$. In the case where masses and concentrations are known, the results are almost equivalent to the case with known matter field in the last section, giving $\sigma_{\fnl} \simeq 1$. This is possible due to the low shot noise in AbacusPNG. Unlike the case where the mass field is known, in the reconstructed mass field analysis presented in this section the $\fnl$ sensitivity somewhat depends on the $\fnl$ value. This may be because $\fnl$ affects the signal-to-noise on the largest scales and thus affects how well we can reconstruct the matter field, which affects the amount of sample variance cancellation. 

We do not compare to a classical mass-binned halo bias analysis without known matter field (the analysis in Fig. \ref{fig:sigma_fnl_fisher_mcmc} assumed that the matter field was known). This is because we had difficulty getting such an analysis to converge, due to the large number of free bias and noise parameters. Without a noise-free matter field for reference, auto- and cross-noises are poorly constrained by the data. This is an advantage of our formalism: By reducing the sum of fields to only two, there are much less free parameters that need to be fit by the MCMC.

\begin{table}[tbh!]

\centering
\begin{minipage}{0.45\linewidth}
\centering
\begin{tabular}{|c|c|c|c|c|}
sim, $f_{NL}$&\multicolumn{2}{c|}{$\hat{f}_{NL}$} & \multicolumn{2}{c|}{$\sigma_{\hat{f}_{NL}}$} \\
\hline
PNG -100   & -98.5 & -101.3 & 4.8 & 3.7  \\
PNG -30    &-20.2 & -30.3 & 6.3 & 4.5    \\
PNG 0      &11.0 & -2.9 & 7.5 & 5.2    \\
PNG +30    & 37.3 & 22.7 & 8.2 & 5.8    \\
PNG +100   & 100.4 & 78.6 & 10.3 & 6.5
   \\
\hline
$k_{max}$ & 0.015 & 0.03 & 0.015 & 0.03 \\
\end{tabular}
\end{minipage}
\hfill
\begin{minipage}{0.45\linewidth}
\centering
\begin{tabular}{|c|c|c|c|c|}
sim, $f_{NL}$&\multicolumn{2}{c|}{$\hat{f}_{NL}$} & \multicolumn{2}{c|}{$\sigma_{\hat{f}_{NL}}$} \\
\hline
PNG -100 & -101.8 & -98.1 & 3.7 & 1.7 \\   
PNG -30  & -30.3 & -29.6 & 1.5 & 1.0 \\
PNG 0    & 0.0 & -0.8 & 1.4 & 1.1 \\
PNG +30  & 28.4 & 26.3 & 2.0 & 1.3  \\
PNG +100 & 95.7 & 90.4 & 3.9 & 2.1 \\
\hline
$k_{max}$ & 0.015 & 0.03 & 0.015 & 0.03 \\
\end{tabular}
\end{minipage}
\caption{Results of Sec. \ref{sec:results_massunknown}. MCMC estimation of $f_{NL}$ on five sets of AbacusPNG simulations with true values of $f_{NL} = \{-30,-100,0,30,100\}$ with the reconstruction of the large-scale matter field $\delta^L_m$ and NN $\pi$-field trained on mass- (left) and mass-concentration (right) binned halo catalogs. The results are averaged over two simulations in each PNG subset.}
\label{tab:fnl_mcmc_2field}
\end{table}

\section{Conclusion}
\label{sec:conclusion}

This paper extends our previous work on neural network enhanced local primordial non-Gaussianity estimation \cite{Giri_2023} from the observation of matter to the more realistic observation of halos, while keeping the crucial property of robustness to non-linear physics of the original method. We introduced a novel machine learning based two-field formalism and applied it to a simple neural network setup which processes local halo population statistics (rather than individual halo positions and properties). We first showed that N halo fields can be combined in only two fields with equivalent sensitivity to $f^{loc}_{NL}$. We then showed that these fields have a natural physical interpretation as a measurement of the local perturbation amplitude $\sigma_8^{loc}$ and of the local large-scale matter density $\delta_m^{loc}$. This suggests that if we can train two neural networks to estimate these fields with maximal statistical sensitivity, the resulting fields will be the optimal fields to base an $\fnl$ estimate on. 

We then demonstrated the approach by analyzing AbacusPNG halo catalogs, comparing our new method with the traditional mass-binned halo analysis. We showed how additional halo features can be seamlessly incorporated into the analysis and increase the signal-to-noise of the analysis. In agreement with previous analyses, we found that halo concentrations can improve the $f_{NL}$ sensitivity by a factor of a few.

The present work is focused on the development of the two-field formalism, and uses only a very simple machine learning model. In forthcoming work, we will replace the CNN acting on halo population statistics with a model that directly processes individual halos and their properties (position, velocity, etc.). This will avoid information loss by mass and concentration binning, and extract information from the precise relative positions of halos and sub-halos or galaxies. A suitable architecture for this approach is a Graph Neural Network that processes the local halo distribution. The power of graph neural networks to constrain cosmological parameters was recently shown in \cite{Villanueva_Domingo_2022, shao2022robust}. Preliminary work by the authors indicates that GNNs can indeed outperform the halo-concentration histogram in measuring $\sigma_8$ at least in the CAMELS simulations \cite{CAMELS:2020cof}. As we have shown here, if $\sigma_8$ can be tightened by the model, this translates directly into tighter $\fnl$ constraints. Another important question to address is the observability of halo features. In this work, we assumed that both halo masses and halo concentrations are directly accessible. In a more realistic scenario, however, we need to take into account the uncertainty in the halo mass estimation and consider directly observable features that can trace halo concentrations (see \cite{Sullivan:2023qjr}).

Because of the simplicity of our halo data in the present work, we cannot yet comprehensively answer what improvement factor neural networks can offer over scale-dependent bias in a realistic experiment. The present paper lays the foundation to answering this question, and upcoming work will explore more powerful machine learning models in the context of our formalism.

\section{Acknowledgements}
We appreciate fruitful discussions with Neal Dalal, Utkarsh Giri and Andrej Obuljen. M.M. acknowledges the support by the U.S. Department of Energy, Office of Science, Office of High Energy Physics under Award Number DE-SC-0017647, and by NSF grant 2307109. K.M.S. was supported by an NSERC Discovery Grant and a CIFAR fellowship. 
Y.K. and M.M. are grateful for the hospitality of Perimeter Institute where a part of this work was done. Research at Perimeter Institute is supported in part by the Government of Canada through the Department of Innovation, Science and Economic Development Canada and by the Province of Ontario through the Ministry of Colleges and Universities. We have extensively used the following python libraries: \texttt{pytorch}\cite{paszke2019pytorchimperativestylehighperformance}, \texttt{numpy}\cite{harris2020array}, \texttt{scipy}\cite{2020SciPy-NMeth}, \texttt{jax}\cite{jax2018github}, \texttt{matplotlib}\cite{Hunter:2007}, \texttt{getdist}\cite{lewis2019getdistpythonpackageanalysing}, \texttt{numba}\cite{10.1145/2833157.2833162}, \texttt{emcee}\cite{Foreman_Mackey_2013}, \texttt{pylians}\cite{Pylians}, \texttt{sympy}\cite{10.7717/peerj-cs.103}, and \texttt{class}\cite{Diego_Blas_2011}.

\printbibliography

@article{Sullivan:2023qjr,
    author = "Sullivan, James M. and Prijon, Tijan and Seljak, Uros",
    title = "{Learning to concentrate: multi-tracer forecasts on local primordial non-Gaussianity with machine-learned~bias}",
    eprint = "2303.08901",
    archivePrefix = "arXiv",
    primaryClass = "astro-ph.CO",
    doi = "10.1088/1475-7516/2023/08/004",
    journal = "JCAP",
    volume = "08",
    pages = "004",
    year = "2023"
}

@article{CAMELS:2020cof,
    author = "Villaescusa-Navarro, Francisco and others",
    collaboration = "CAMELS",
    title = "{The CAMELS project: Cosmology and Astrophysics with MachinE Learning Simulations}",
    eprint = "2010.00619",
    archivePrefix = "arXiv",
    primaryClass = "astro-ph.CO",
    doi = "10.3847/1538-4357/abf7ba",
    journal = "Astrophys. J.",
    volume = "915",
    pages = "71",
    year = "2021"
}

@article{Andrews_2023,
   title={Bayesian field-level inference of primordial non-Gaussianity using next-generation galaxy surveys},
   volume={520},
   ISSN={1365-2966},
   url={http://dx.doi.org/10.1093/mnras/stad432},
   DOI={10.1093/mnras/stad432},
   number={4},
   journal={Monthly Notices of the Royal Astronomical Society},
   publisher={Oxford University Press (OUP)},
   author={Andrews, Adam and Jasche, Jens and Lavaux, Guilhem and Schmidt, Fabian},
   year={2023},
   month=feb, pages={5746–5763} }

@article{Biagetti_2021,
   title={The persistence of large scale structures. Part I. Primordial non-Gaussianity},
   volume={2021},
   ISSN={1475-7516},
   url={http://dx.doi.org/10.1088/1475-7516/2021/04/061},
   DOI={10.1088/1475-7516/2021/04/061},
   number={04},
   journal={Journal of Cosmology and Astroparticle Physics},
   publisher={IOP Publishing},
   author={Biagetti, Matteo and Cole, Alex and Shiu, Gary},
   year={2021},
   month=apr, pages={061} }

@article{Seljak_2009,
   title={Extracting Primordial Non-Gaussianity without Cosmic Variance},
   volume={102},
   ISSN={1079-7114},
   url={http://dx.doi.org/10.1103/PhysRevLett.102.021302},
   DOI={10.1103/physrevlett.102.021302},
   number={2},
   journal={Physical Review Letters},
   publisher={American Physical Society (APS)},
   author={Seljak, Uroš},
   year={2009},
   month=jan }

@article{Castorina_2018,
   title={Primordial Non-Gaussianities and Zero-Bias Tracers of the Large-Scale Structure},
   volume={121},
   ISSN={1079-7114},
   url={http://dx.doi.org/10.1103/PhysRevLett.121.101301},
   DOI={10.1103/physrevlett.121.101301},
   number={10},
   journal={Physical Review Letters},
   publisher={American Physical Society (APS)},
   author={Castorina, Emanuele and Feng, Yu and Seljak, Uroš and Villaescusa-Navarro, Francisco},
   year={2018},
   month=sep }

@article{Goldstein_2022,
  title = {Squeezing ${f}_{\mathrm{NL}}$ out of the matter bispectrum with consistency relations},
  author = {Goldstein, Samuel and Esposito, Angelo and Philcox, Oliver H. E. and Hui, Lam and Hill, J. Colin and Scoccimarro, Roman and Abitbol, Maximilian H.},
  journal = {Phys. Rev. D},
  volume = {106},
  issue = {12},
  pages = {123525},
  numpages = {17},
  year = {2022},
  month = {Dec},
  publisher = {American Physical Society},
  doi = {10.1103/PhysRevD.106.123525},
  url = {https://link.aps.org/doi/10.1103/PhysRevD.106.123525}
}

@misc{li2022pmwddifferentiablecosmologicalparticlemesh,
      title={pmwd: A Differentiable Cosmological Particle-Mesh $N$-body Library}, 
      author={Yin Li and Libin Lu and Chirag Modi and Drew Jamieson and Yucheng Zhang and Yu Feng and Wenda Zhou and Ngai Pok Kwan and François Lanusse and Leslie Greengard},
      year={2022},
      eprint={2211.09958},
      archivePrefix={arXiv},
      primaryClass={astro-ph.IM},
      url={https://arxiv.org/abs/2211.09958}, 
}

@article{Alvarez:2014vva,
    author = "Alvarez, Marcelo and others",
    title = "{Testing Inflation with Large Scale Structure: Connecting Hopes with Reality}",
    eprint = "1412.4671",
    archivePrefix = "arXiv",
    primaryClass = "astro-ph.CO",
    month = "12",
    year = "2014"
}

@article{Dalal:2007cu,
	Archiveprefix = {arXiv},
	Author = {Dalal, Neal and Dore, Olivier and Huterer, Dragan and Shirokov, Alexander},
	Doi = {10.1103/PhysRevD.77.123514},
	Eprint = {0710.4560},
	Journal = {Phys. Rev.},
	Pages = {123514},
	Primaryclass = {astro-ph},
	Slaccitation = {%%CITATION = ARXIV:0710.4560;%%},
	Title = {{The imprints of primordial non-gaussianities on large-scale structure: scale dependent bias and abundance of virialized objects}},
	Volume = {D77},
	Year = {2008}}

@article{Giri:2023mpg,
    author = {Giri, Utkarsh and M\"unchmeyer, Moritz and Smith, Kendrick M.},
    title = "{Constraining $f_{NL}$ using the Large-Scale Modulation of Small-Scale Statistics}",
    eprint = "2305.03070",
    archivePrefix = "arXiv",
    primaryClass = "astro-ph.CO",
    month = "5",
    year = "2023"
}

@article{Giri_2023,
   title={Robust neural network-enhanced estimation of local primordial non-Gaussianity},
   volume={107},
   ISSN={2470-0029},
   url={http://dx.doi.org/10.1103/PhysRevD.107.L061301},
   DOI={10.1103/physrevd.107.l061301},
   number={6},
   journal={Physical Review D},
   publisher={American Physical Society (APS)},
   author={Giri, Utkarsh and Münchmeyer, Moritz and Smith, Kendrick M.},
   year={2023},
   month=mar }

@article{Villanueva_Domingo_2022,
   title={Learning Cosmology and Clustering with Cosmic Graphs},
   volume={937},
   ISSN={1538-4357},
   url={http://dx.doi.org/10.3847/1538-4357/ac8930},
   DOI={10.3847/1538-4357/ac8930},
   number={2},
   journal={The Astrophysical Journal},
   publisher={American Astronomical Society},
   author={Villanueva-Domingo, Pablo and Villaescusa-Navarro, Francisco},
   year={2022},
   month=oct, pages={115} }

@misc{shao2022robust,
      title={Robust field-level inference with dark matter halos}, 
      author={Helen Shao and Francisco Villaescusa-Navarro and Pablo Villanueva-Domingo and Romain Teyssier and Lehman H. Garrison and Marco Gatti and Derek Inman and Yueying Ni and Ulrich P. Steinwandel and Mihir Kulkarni and Eli Visbal and Greg L. Bryan and Daniel Angles-Alcazar and Tiago Castro and Elena Hernandez-Martinez and Klaus Dolag},
      year={2022},
      eprint={2209.06843},
      archivePrefix={arXiv},
      primaryClass={astro-ph.CO}
}

@misc{hadzhiyska2024abacuspng,
      title={AbacusPNG: A modest set of simulations of local-type primordial non-Gaussianity in the DESI era}, 
      author={Boryana Hadzhiyska and Lehman Garrison and Daniel J. Eisenstein and Simone Ferraro},
      year={2024},
      eprint={2402.10881},
      archivePrefix={arXiv},
      primaryClass={astro-ph.CO}
}

@ARTICLE{10.1093/mnras/stab2484,
    author = {Maksimova, Nina A and Garrison, Lehman H and Eisenstein, Daniel J and Hadzhiyska, Boryana and Bose, Sownak and Satterthwaite, Thomas P},
    title = "{AbacusSummit: a massive set of high-accuracy, high-resolution N-body simulations}",
    journal = {Monthly Notices of the Royal Astronomical Society},
    volume = {508},
    number = {3},
    pages = {4017-4037},
    year = {2021},
    month = {09},
    issn = {0035-8711},
    doi = {10.1093/mnras/stab2484},
    url = {https://doi.org/10.1093/mnras/stab2484},
    eprint = {https://academic.oup.com/mnras/article-pdf/508/3/4017/40811763/stab2484.pdf},
}

@ARTICLE{10.1093/mnras/stab2482,
    author = {Garrison, Lehman H and Eisenstein, Daniel J and Ferrer, Douglas and Maksimova, Nina A and Pinto, Philip A},
    title = "{The abacus cosmological N-body code}",
    journal = {Monthly Notices of the Royal Astronomical Society},
    volume = {508},
    number = {1},
    pages = {575-596},
    year = {2021},
    month = {09},
    issn = {0035-8711},
    doi = {10.1093/mnras/stab2482},
    url = {https://doi.org/10.1093/mnras/stab2482},
    eprint = {https://academic.oup.com/mnras/article-pdf/508/1/575/40458823/stab2482.pdf},
}

@ARTICLE{10.1093/mnras/stab2980,
    author = {Hadzhiyska, Boryana and Eisenstein, Daniel and Bose, Sownak and Garrison, Lehman H and Maksimova, Nina},
    title = "{CompaSO: A new halo finder for competitive assignment to spherical overdensities}",
    journal = {Monthly Notices of the Royal Astronomical Society},
    year = {2021},
    month = {10},
    issn = {0035-8711},
    doi = {10.1093/mnras/stab2980},
    url = {https://doi.org/10.1093/mnras/stab2980},
    note = {stab2980},
    eprint = {https://academic.oup.com/mnras/advance-article-pdf/doi/10.1093/mnras/stab2980/40751402/stab2980.pdf},
}

@misc{rocher2024desi,
      title={The DESI One-Percent survey: exploring the Halo Occupation Distribution of Emission Line Galaxies with AbacusSummit simulations}, 
      author={Antoine Rocher and Vanina Ruhlmann-Kleider and Etienne Burtin and Sihan Yuan and Arnaud de Mattia and Ashley J. Ross and Jessica Aguilar and Steven Ahlen and Shadab Alam and Davide Bianchi and David Brooks and Shaun Cole and Kyle Dawson and Axel de la Macorra and Peter Doel and Daniel J. Eisenstein and Kevin Fanning and Jaime E. Forero-Romero and Lehman H. Garrison and Satya Gontcho A Gontcho and Violeta Gonzalez-Perez and Julien Guy and Boryana Hadzhiyska and ChangHoon Hahn and Klaus Honscheid and Theodore Kisner and Martin Landriau and James Lasker and Michael E. Levi and Marc Manera and Aaron Meisner and Ramon Miquel and John Moustakas and Eva-Maria Mueller and Jeffrey A. Newman and Jundan Nie and Will J. Percival and Claire Poppett and Fei Qin and Graziano Rossi and Lado Samushia and Eusebio Sanchez and David Schlegel and Michael Schubnell and Hee-Jong Seo and Gregory Tarlé and Mariana Vargas-Magaña and Benjamin A. Weaver and Jiaxi Yu and Hanyu Zhang and Zheng Zheng and Zhimin Zhou and Hu Zou},
      year={2024},
      eprint={2306.06319},
      archivePrefix={arXiv},
      primaryClass={astro-ph.CO}
}

@article{Lazeyras_2023,
   title={Assembly bias in the local PNG halo bias and its implication for f
               NL constraints},
   volume={2023},
   ISSN={1475-7516},
   url={http://dx.doi.org/10.1088/1475-7516/2023/01/023},
   DOI={10.1088/1475-7516/2023/01/023},
   number={01},
   journal={Journal of Cosmology and Astroparticle Physics},
   publisher={IOP Publishing},
   author={Lazeyras, Titouan and Barreira, Alexandre and Schmidt, Fabian and Desjacques, Vincent},
   year={2023},
   month=jan, pages={023} }

@article{Villaescusa-Navarro:2019bje,
    author = "Villaescusa-Navarro, Francisco and others",
    title = "{The Quijote simulations}",
    eprint = "1909.05273",
    archivePrefix = "arXiv",
    primaryClass = "astro-ph.CO",
    doi = "10.3847/1538-4365/ab9d82",
    journal = "Astrophys. J. Suppl.",
    volume = "250",
    number = "1",
    pages = "2",
    year = "2020"
}

@article{Ferraro:2014jba,
    author = "Ferraro, Simone and Smith, Kendrick M.",
    title = "{Using large scale structure to measure $f_{NL}, g_{NL}$ and $τ_{NL}$}",
    eprint = "1408.3126",
    archivePrefix = "arXiv",
    primaryClass = "astro-ph.CO",
    doi = "10.1103/PhysRevD.91.043506",
    journal = "Phys. Rev. D",
    volume = "91",
    number = "4",
    pages = "043506",
    year = "2015"
}

@article{Giri_2022,
   title={Exploring KSZ velocity reconstruction with N-body simulations and the halo model},
   volume={2022},
   ISSN={1475-7516},
   url={http://dx.doi.org/10.1088/1475-7516/2022/09/028},
   DOI={10.1088/1475-7516/2022/09/028},
   number={09},
   journal={Journal of Cosmology and Astroparticle Physics},
   publisher={IOP Publishing},
   author={Giri, Utkarsh and Smith, Kendrick M.},
   year={2022},
   month=sep, pages={028} }

@misc{smith2018ksztomographybispectrum,
      title={KSZ tomography and the bispectrum}, 
      author={Kendrick M. Smith and Mathew S. Madhavacheril and Moritz Münchmeyer and Simone Ferraro and Utkarsh Giri and Matthew C. Johnson},
      year={2018},
      eprint={1810.13423},
      archivePrefix={arXiv},
      primaryClass={astro-ph.CO},
      url={https://arxiv.org/abs/1810.13423}, 
}

@article{M_nchmeyer_2019,
   title={Constraining local non-Gaussianities with kinetic Sunyaev-Zel’dovich tomography},
   volume={100},
   ISSN={2470-0029},
   url={http://dx.doi.org/10.1103/PhysRevD.100.083508},
   DOI={10.1103/physrevd.100.083508},
   number={8},
   journal={Physical Review D},
   publisher={American Physical Society (APS)},
   author={Münchmeyer, Moritz and Madhavacheril, Mathew S. and Ferraro, Simone and Johnson, Matthew C. and Smith, Kendrick M.},
   year={2019},
   month=oct }

@article{Schmittfull_2018,
   title={Parameter constraints from cross-correlation of CMB lensing with galaxy clustering},
   volume={97},
   ISSN={2470-0029},
   url={http://dx.doi.org/10.1103/PhysRevD.97.123540},
   DOI={10.1103/physrevd.97.123540},
   number={12},
   journal={Physical Review D},
   publisher={American Physical Society (APS)},
   author={Schmittfull, Marcel and Seljak, Uroš},
   year={2018},
   month=jun }

@article{Cayuso_2023,
   title={Velocity reconstruction with the cosmic microwave background and galaxy surveys},
   volume={2023},
   ISSN={1475-7516},
   url={http://dx.doi.org/10.1088/1475-7516/2023/02/051},
   DOI={10.1088/1475-7516/2023/02/051},
   number={02},
   journal={Journal of Cosmology and Astroparticle Physics},
   publisher={IOP Publishing},
   author={Cayuso, Juan and Bloch, Richard and Hotinli, Selim C. and Johnson, Matthew C. and McCarthy, Fiona},
   year={2023},
   month=feb, pages={051} }

@article{Deutsch_2018,
   title={Reconstruction of the remote dipole and quadrupole fields from the kinetic Sunyaev Zel’dovich and polarized Sunyaev Zel’dovich effects},
   volume={98},
   ISSN={2470-0029},
   url={http://dx.doi.org/10.1103/PhysRevD.98.123501},
   DOI={10.1103/physrevd.98.123501},
   number={12},
   journal={Physical Review D},
   publisher={American Physical Society (APS)},
   author={Deutsch, Anne-Sylvie and Dimastrogiovanni, Emanuela and Johnson, Matthew C. and Münchmeyer, Moritz and Terrana, Alexandra},
   year={2018},
   month=dec }

@article{Contreras_2023,
   title={Maximum likelihood kinetic Sunyaev-Zel’dovich velocity reconstruction},
   volume={107},
   ISSN={2470-0029},
   url={http://dx.doi.org/10.1103/PhysRevD.107.023521},
   DOI={10.1103/physrevd.107.023521},
   number={2},
   journal={Physical Review D},
   publisher={American Physical Society (APS)},
   author={Contreras, Dagoberto and McCarthy, Fiona and Johnson, Matthew C.},
   year={2023},
   month=jan }

@misc{kvasiuk2024autodiff,
      title={An Auto-Differentiable Likelihood Pipeline for the Cross-Correlation of CMB and Large-Scale Structure due to the Kinetic Sunyaev-Zeldovich Effect}, 
      author={Yurii Kvasiuk and Moritz Münchmeyer},
      year={2024},
      eprint={2305.08903},
      archivePrefix={arXiv},
      primaryClass={astro-ph.CO},
      url={https://arxiv.org/abs/2305.08903}, 
}

@misc{paszke2019pytorchimperativestylehighperformance,
      title={PyTorch: An Imperative Style, High-Performance Deep Learning Library}, 
      author={Adam Paszke and Sam Gross and Francisco Massa and Adam Lerer and James Bradbury and Gregory Chanan and Trevor Killeen and Zeming Lin and Natalia Gimelshein and Luca Antiga and Alban Desmaison and Andreas Köpf and Edward Yang and Zach DeVito and Martin Raison and Alykhan Tejani and Sasank Chilamkurthy and Benoit Steiner and Lu Fang and Junjie Bai and Soumith Chintala},
      year={2019},
      eprint={1912.01703},
      archivePrefix={arXiv},
      primaryClass={cs.LG},
      url={https://arxiv.org/abs/1912.01703}, 
}

@Article{         harris2020array,
 title         = {Array programming with {NumPy}},
 author        = {Charles R. Harris and K. Jarrod Millman and St{\'{e}}fan J.
                 van der Walt and Ralf Gommers and Pauli Virtanen and David
                 Cournapeau and Eric Wieser and Julian Taylor and Sebastian
                 Berg and Nathaniel J. Smith and Robert Kern and Matti Picus
                 and Stephan Hoyer and Marten H. van Kerkwijk and Matthew
                 Brett and Allan Haldane and Jaime Fern{\'{a}}ndez del
                 R{\'{i}}o and Mark Wiebe and Pearu Peterson and Pierre
                 G{\'{e}}rard-Marchant and Kevin Sheppard and Tyler Reddy and
                 Warren Weckesser and Hameer Abbasi and Christoph Gohlke and
                 Travis E. Oliphant},
 year          = {2020},
 month         = sep,
 journal       = {Nature},
 volume        = {585},
 number        = {7825},
 pages         = {357--362},
 doi           = {10.1038/s41586-020-2649-2},
 publisher     = {Springer Science and Business Media {LLC}},
 url           = {https://doi.org/10.1038/s41586-020-2649-2}
}

@ARTICLE{2020SciPy-NMeth,
  author  = {Virtanen, Pauli and Gommers, Ralf and Oliphant, Travis E. and
            Haberland, Matt and Reddy, Tyler and Cournapeau, David and
            Burovski, Evgeni and Peterson, Pearu and Weckesser, Warren and
            Bright, Jonathan and {van der Walt}, St{\'e}fan J. and
            Brett, Matthew and Wilson, Joshua and Millman, K. Jarrod and
            Mayorov, Nikolay and Nelson, Andrew R. J. and Jones, Eric and
            Kern, Robert and Larson, Eric and Carey, C J and
            Polat, {\.I}lhan and Feng, Yu and Moore, Eric W. and
            {VanderPlas}, Jake and Laxalde, Denis and Perktold, Josef and
            Cimrman, Robert and Henriksen, Ian and Quintero, E. A. and
            Harris, Charles R. and Archibald, Anne M. and
            Ribeiro, Ant{\^o}nio H. and Pedregosa, Fabian and
            {van Mulbregt}, Paul and {SciPy 1.0 Contributors}},
  title   = {{{SciPy} 1.0: Fundamental Algorithms for Scientific
            Computing in Python}},
  journal = {Nature Methods},
  year    = {2020},
  volume  = {17},
  pages   = {261--272},
  adsurl  = {https://rdcu.be/b08Wh},
  doi     = {10.1038/s41592-019-0686-2},
}

@software{jax2018github,
  author = {James Bradbury and Roy Frostig and Peter Hawkins and Matthew James Johnson and Chris Leary and Dougal Maclaurin and George Necula and Adam Paszke and Jake Vander{P}las and Skye Wanderman-{M}ilne and Qiao Zhang},
  title = {{JAX}: composable transformations of {P}ython+{N}um{P}y programs},
  url = {http://github.com/google/jax},
  version = {0.3.13},
  year = {2018},
}

@Article{Hunter:2007,
  Author    = {Hunter, J. D.},
  Title     = {Matplotlib: A 2D graphics environment},
  Journal   = {Computing in Science \& Engineering},
  Volume    = {9},
  Number    = {3},
  Pages     = {90--95},
  publisher = {IEEE COMPUTER SOC},
  doi       = {10.1109/MCSE.2007.55},
  year      = 2007
}

@misc{lewis2019getdistpythonpackageanalysing,
      title={GetDist: a Python package for analysing Monte Carlo samples}, 
      author={Antony Lewis},
      year={2019},
      eprint={1910.13970},
      archivePrefix={arXiv},
      primaryClass={astro-ph.IM},
      url={https://arxiv.org/abs/1910.13970}, 
}

@inproceedings{10.1145/2833157.2833162, author = {Lam, Siu Kwan and Pitrou, Antoine and Seibert, Stanley}, title = {Numba: a LLVM-based Python JIT compiler}, year = {2015}, isbn = {9781450340052}, publisher = {Association for Computing Machinery}, address = {New York, NY, USA}, url = {https://doi.org/10.1145/2833157.2833162}, doi = {10.1145/2833157.2833162}, booktitle = {Proceedings of the Second Workshop on the LLVM Compiler Infrastructure in HPC}, articleno = {7}, numpages = {6}, location = {Austin, Texas}, series = {LLVM '15} }

@article{Foreman_Mackey_2013,
   title={<tt>emcee</tt>: The MCMC Hammer},
   volume={125},
   ISSN={1538-3873},
   url={http://dx.doi.org/10.1086/670067},
   DOI={10.1086/670067},
   number={925},
   journal={Publications of the Astronomical Society of the Pacific},
   publisher={IOP Publishing},
   author={Foreman-Mackey, Daniel and Hogg, David W. and Lang, Dustin and Goodman, Jonathan},
   year={2013},
   month=mar, pages={306–312} }

@MISC{Pylians,
    author = {{Villaescusa-Navarro}, Francisco},
    title = "{Pylians: Python libraries for the analysis of numerical simulations}",
    howpublished = {Astrophysics Source Code Library, record ascl:1811.008},
    year = 2018,
    month = nov,
    eid = {ascl:1811.008},
    pages = {ascl:1811.008},
    archivePrefix = {ascl},
    eprint = {1811.008},
    adsurl = {https://ui.adsabs.harvard.edu/abs/2018ascl.soft11008V}
}

@article{10.7717/peerj-cs.103,
     title = {SymPy: symbolic computing in Python},
     author = {Meurer, Aaron and Smith, Christopher P. and Paprocki, Mateusz and \v{C}ert\'{i}k, Ond\v{r}ej and Kirpichev, Sergey B. and Rocklin, Matthew and Kumar, AMiT and Ivanov, Sergiu and Moore, Jason K. and Singh, Sartaj and Rathnayake, Thilina and Vig, Sean and Granger, Brian E. and Muller, Richard P. and Bonazzi, Francesco and Gupta, Harsh and Vats, Shivam and Johansson, Fredrik and Pedregosa, Fabian and Curry, Matthew J. and Terrel, Andy R. and Rou\v{c}ka, \v{S}t\v{e}p\'{a}n and Saboo, Ashutosh and Fernando, Isuru and Kulal, Sumith and Cimrman, Robert and Scopatz, Anthony},
     year = 2017,
     month = jan,
     volume = 3,
     pages = {e103},
     journal = {PeerJ Computer Science},
     issn = {2376-5992},
     url = {https://doi.org/10.7717/peerj-cs.103},
     doi = {10.7717/peerj-cs.103}
    }

@article{Diego_Blas_2011,
   title={The Cosmic Linear Anisotropy Solving System (CLASS).
 Part II: Approximation schemes},
   volume={2011},
   ISSN={1475-7516},
   url={http://dx.doi.org/10.1088/1475-7516/2011/07/034},
   DOI={10.1088/1475-7516/2011/07/034},
   number={07},
   journal={Journal of Cosmology and Astroparticle Physics},
   publisher={IOP Publishing},
   author={Diego Blas and Julien Lesgourgues and Thomas Tram},
   year={2011},
   month=jul, pages={034–034} }

@article{Barreira_2020,
   title={Galaxy bias and primordial non-Gaussianity: insights from galaxy formation simulations with IllustrisTNG},
   volume={2020},
   ISSN={1475-7516},
   url={http://dx.doi.org/10.1088/1475-7516/2020/12/013},
   DOI={10.1088/1475-7516/2020/12/013},
   number={12},
   journal={Journal of Cosmology and Astroparticle Physics},
   publisher={IOP Publishing},
   author={Barreira, Alexandre and Cabass, Giovanni and Schmidt, Fabian and Pillepich, Annalisa and Nelson, Dylan},
   year={2020},
   month=dec, pages={013–013} }
\clearpage
\appendix
\section{Fisher information on $\fnl$ with known matter field}
\label{app:fisherwithm}

In this appendix, we derive the Fisher information in the (experimentally unrealistic) case that the matter field is exactly known. We also show that a single $\pi$ field, suitably defined, includes the complete Fisher information. In the following appendix we consider the realistic case with unknown matter field. 

\subsection{Simple expression for the Fisher information}
For the covariance
\begin{equation}
    \mathbf{C} = \begin{bmatrix}
     P_{mm}(k) & b_iP_{mm}(k) \\
    b_jP_{mm}(k) & b_ib_jP_{mm} + N_{ij} \\
\end{bmatrix}
\end{equation}
where $b$ is an $n$-component bias vector $b_{i} = b^{G}_{i} + 2b^{NG}_{i}\frac{f_{NL}}{\alpha}$, $N_{ij}$ is a correlated noise power spectrum and $P_{mm}(k)$ is a large-scale matter power spectrum, we need to calculate
\begin{equation}
    \mathcal{F}_{ab} = \mathrm{Tr}\left[C^{-1}C_{,a}C^{-1}C_{,b}\right]
\end{equation}
Let's redefine factor out matter power spectrum and redefine b and N so that the covariance can be written in the following block form

\begin{equation}
C = P_{mm}\begin{bmatrix}
    1 & b^T \\
    b & bb^T + N
\end{bmatrix}
\end{equation}
For the general block matrix (assuming corresponding inverses exist) we have:
\begin{equation}
   \begin{bmatrix}
    A_{11} & A_{12} \\
    A_{21} & A_{22}
\end{bmatrix}^{-1} = 
\begin{bmatrix}
B_{1}^{-1} & -B^{-1}_{1}A_{12}A^{-1}_{22} \\
-A^{-1}_{22}A_{21}B^{-1}_1 & B^{-1}_2
\end{bmatrix} 
\end{equation}
Where $B_1$ and $B_2$ are Schur complements: $B_1 = A_{11} - A_{12}A^{-1}_{22}A_{21}$ and 
$B_2 = A_{22} - A_{21}A^{-1}_{11}A_{12}$
For the matrix $C$, we have $B_1 = 1 - b^T(N+bb^T)^{-1}b$. 
Sherman-Morrison formula tells us that 
\begin{equation}
    (A+bc^T) = A^{-1} - \frac{A^{-1}bc^TA^{-1}}{1+c^TA^{-1}b}
\end{equation}
That leads to 
\begin{equation}
    B_1 = \frac{1}{1+b^TN^{-1}b}
\end{equation}
Next,
\begin{align}
    -B^{-1}_{1}A_{12}A^{-1}_{22} &= -(1+b^TN^{-1}b)(b^TN^{-1}-\frac{b^TN^{-1}bb^TN^{-1}}{1+b^TN^{-1}b}) \\
    &= -(1+b^TN^{-1}b)(1-\frac{b^TN^{-1}b}{1+b^TN^{-1}b})b^TN^{-1} \\
    &= -b^TN^{-1}
\end{align}
Finally, 
\begin{equation}
B_2 = bb^T+N-bb^T = N
\end{equation}
Hence, the inverse of covariance is as follows:
\begin{equation}
C^{-1} = \begin{bmatrix}
    1+b^TN^{-1}b & -b^TN^{-1} \\
    -N^{-1}b & N^{-1}
\end{bmatrix}   
\end{equation}
Let's define $\frac{\partial b}{\partial f_{NL}} = d$ (It's just a non-Gaussian bias vector up to a constant in the conventional notation) Then $\frac{\partial C}{\partial f_{NL}}$ is as follows:

\begin{equation}
\frac{\partial C}{\partial f_{NL}} = \begin{bmatrix}
    0 & d^T \\
    d & db^T + bd^T
\end{bmatrix}   
\end{equation}
Multiplying two block matrices, one can find
\begin{equation}
C^{-1}\frac{\partial C}{\partial f_{NL}} = \begin{bmatrix}
    -b^TN^{-1}d & (1+b^TN^{-1}b)d^T -b^TN^{-1}(db^T + bd^T) \\
    N^{-1}d & -N^{-1}bd^T + N^{-1}(db^T + bd^T)
\end{bmatrix}   
\end{equation}
Then 
\begin{align}
    \mathrm{Tr}\left[C^{-1}C_{,f_{NL}}C^{-1}C_{,f_{NL}}\right] & = 
    (b^TN^{-1}d)^2 +2\left((d^TN^{-1}d-(b^TN^{-1}d)^2)\right) \\
    & + \mathrm{Tr}\left[(-N^{-1}bd^T + N^{-1}(db^T + bd^T))^2\right] \\
    & = 2 d^TN^{-1}d
\end{align}
So that (back to the original notation)
\begin{equation}
    \mathcal{F}_{f_{NL}f_{NL}} = 8\frac{P_{mm}}{\alpha^2}b^{NG}_i(N^{-1})_{ij}b^{NG}_j
\end{equation}

Similarly, one can find the full Fisher matrix. Mainly,
\begin{align}
\mathcal{F}_{f_{NL}b_j^G} = 4\frac{P_{mm}}{\alpha}b^{NG}_{i}(N^{-1})_{ij} \\
\mathcal{F}_{b_i^Gb_j^G} = 2P_{mm}(N^{-1})_{ij}
\end{align}
One also notes that it takes block-diagonal form with an upper block as described above and a lower block that corresponds to derivatives wrt noise. Noticeably, mixed bias-noise components are zero allowing to only invert an upper block to get marginalized error bound.
\begin{equation}
    F_{ab} = \frac{1}{2}\sum_{k}\mathcal{F}_{ab}(k) =  \begin{pmatrix}
    c_{00}b^{NG}_i(N^{-1})_{ij}b^{NG}_j & c_{01} b^{NG}_{i}(N^{-1})_{ij} \\
    c_{01} (N^{-1})_{ij}b^{NG}_{j} & c_{11}(N^{-1})_{ij}
    \end{pmatrix}
\end{equation}
Where $c_{ij}$ are the same as previously defined:
\begin{align}
    c_{00} = \sum_{k}\frac{8P_{mm}(k)}{\alpha^2(k)};\ \ \ c_{01} = \sum_{k}\frac{4P_{mm}(k)}{\alpha(k)};\ \ \ c_{11} = \sum_{k}2P_{mm}(k);
\end{align}
For the inverse, one finds:
\begin{equation}
(F^{-1})_{f_{NL}f_{NL}} = \frac{\mathrm{det}N^{-1}}{\mathrm{det}F} = \frac{c_{11}}{c_{00}c_{11}-c^2_{01}}\frac{1}{b^{NG}_i(N^{-1})_{ij}b^{NG}_j}
\end{equation}

\subsection{Fisher information with a single optimal $\pi$-field}

As we saw in the previous section the Fisher information is given by

\begin{equation}
(F^{-1})_{f_{NL}f_{NL}} = \frac{c_{11}}{c_{00}c_{11}-c^2_{01}}\frac{1}{b^{NG}_i(N^{-1})_{ij}b^{NG}_j}
\end{equation}

This suggests that we can compress the multiple $\pi$ fields into a single field that contains the same amount of information. Define a new single field by
\begin{align}
    \pi' = b_{ng}^T N^{-1} \vec{\pi}
\end{align}
This field has biases
\begin{align}
    b_{g}' = b_{ng}^T N^{-1} b_{g}  \hspace{1cm} b_{ng}' = b_{ng}^T N^{-1} b_{ng}
\end{align}
and noise
\begin{align}
    N' = b_{ng}^T N^{-1} b_{ng} 
\end{align}

This new field has the same bilinear form
\begin{align}
   b_{ng}^T (N')^{-1} b'_{ng} = b_{ng}^T N^{-1} b_{ng} 
\end{align} 
as the original fields and thus the Fisher information is the same. The Fisher information of the new field can be written as
\begin{equation}
(F^{-1})_{f_{NL}f_{NL}} = \frac{c_{11}}{c_{00}c_{11}-c^2_{01}}\frac{(b'_{ng})^2}{N'}
\end{equation}

\section{Fisher information on $\fnl$ with unknown matter field}
\label{app:fishernomatter}
In this section we will construct two fields that contain the same Fisher information on $\fnl$ as a set of arbitrary many $\pi$ fields.

\subsection{Equation for the Fisher information}

In the case where we don't have a matter field, we still can calculate $\mathcal{F}_{f_{NL}f_{NL}}$, however, the expression does not simplyfy as much. For that, we need to calculate the following trace:

\begin{align}
\mathcal{F}_{f_{NL}f_{NL}} &= \mathrm{Tr}\left[\left((db^T+bd^T)(N^{-1}-\frac{N^{-1}bb^TN^{-1}}{1+b^TN^{-1}b})\right)^2\right] \\
&= 2\left[(b^TN^{-1}d)^2\frac{1-b^TN^{-1}b}{(1+b^TN^{-1}b)^2} + (d^TN^{-1}d)\frac{b^TN^{-1}b}{1+b^TN^{-1}b}\right]
\end{align}
We note that it's only dependent on the three bilinear forms: $b^TN^{-1}b$, $d^TN^{-1}b$, and $d^TN^{-1}d$. The bilinear forms are basis-invariant. This equation presents the general expression for the sensitivity to $f_{NL}$ of an analysis containing N linearly-biased tracers with possibly correlated noises. Some partial cases were considered in \cite{Sullivan:2023qjr}, \cite{M_nchmeyer_2019} and \cite{Schmittfull_2018}.

\subsection{Constructing fields with arbitrary biases}

We first want to create a field $\pi^0$ with arbitrary choice of $b^0_g$ and $b^0_{ng}$ by weighting the input fields as
\begin{align}
	\pi^0 = \sum_i w_i \pi_i = w^T \pi
\end{align}
If we have more than two fields $\pi_i$  there are many different choices, so we add the condition that the noise of the resulting field $\pi_0$ should be minimal. We thus have a constraint optimization problem. We minimize
\begin{align}
	\left< \epsilon \epsilon^T \right> = w^T N^{-1} w
\end{align}
with constraints $w^T b_g = b_g^0$ and $w^T b_{ng} = b_{ng}^{0}$.
Given:
\begin{itemize}
  \item Vectors \( w, b_g, b_{ng} \)
  \item Symmetric positive-definite matrix \( N \)
  \item Real numbers \( b_g^0 \) and \( b_{ng}^0 \)
\end{itemize}
we minimize \( w^T N w \) subject to \( w^T b_g = b_g^0 \) and \( w^T b_{ng} = b_{ng}^0 \).
The Lagrangian \( \mathcal{L} \) for this optimization problem is defined as:
\begin{align}
\mathcal{L}(w, \lambda, \mu) = w^T N w + \lambda (w^T b_g - b_g^0) + \mu (w^T b_{ng} - b_{ng}^0)
\end{align}
Taking the gradient of \( \mathcal{L} \) with respect to \( w \) and setting it to zero yields:
\begin{align}
\nabla_w \mathcal{L} = 2Nw + \lambda b_g + \mu b_{ng} = 0
\end{align}
which we can solve for \( w \) as:
\begin{align}
w = -\frac{1}{2} N^{-1} (\lambda b_g + \mu b_{ng})
\end{align}
The constraint equations are:
\begin{align}
w^T b_g = b_g^0 \quad \text{and} \quad w^T b_{ng} = b_{ng}^0
\end{align}
Plugging the solution for $w$ into the constraint equations gives
\begin{align}
\begin{bmatrix}
b_g^T N^{-1} b_g & b_g^T N^{-1} b_{ng} \\
b_{ng}^T N^{-1} b_g & b_{ng}^T N^{-1} b_{ng}
\end{bmatrix}
\begin{bmatrix}
\lambda \\
\mu
\end{bmatrix}
=
\begin{bmatrix}
-2 b_g^0 \\
-2 b_{ng}^0
\end{bmatrix}
\end{align}
The above linear system is solved by (assuming the inverse exist, e.g. the two bias vectors are not co-linear):
\begin{align}
\begin{bmatrix}
\lambda \\
\mu
\end{bmatrix}
=
\frac{1}{\text{det}(A)}
\begin{bmatrix}
b_{ng}^T N^{-1} b_{ng} & -b_g^T N^{-1} b_{ng} \\
-b_{ng}^T N^{-1} b_g & b_g^T N^{-1} b_g
\end{bmatrix}
\begin{bmatrix}
-2 b_g^0 \\
-2 b_{ng}^0
\end{bmatrix}
\end{align}
We then substitute the solution back into $w$.

\subsection{Constructing fields with only Gaussian or only non-Gaussian bias}
\label{sec:2fieldweights}
Based on the intuition that we need to optimally measure both the matter field and the local $\sigma_8$ field we construct two fields with properties 
\begin{itemize}
    \item $\pi_m$ with $b_g=1$ and $b_{ng}=0$
    \item $\pi_\sigma$ with $b_g=0$ and $b_{ng}=1$
\end{itemize}

The solution is

\begin{align}
\label{eq:fieldweights1}
w_m =  \left( \frac{(b_{ng}^T N^{-1} b_{ng}) N^{-1} b_g - (b_{ng}^T N^{-1} b_g) N^{-1} b_{ng}}{(b_g^T N^{-1} b_g) (b_{ng}^T N^{-1} b_{ng}) - (b_g^T N^{-1} b_{ng})^2} \right)
\end{align}

\begin{align}
\label{eq:fieldweights2}
w_\sigma =  \left( \frac{(b_g^T N^{-1} b_g) N^{-1} b_{ng} - (b_g^T N^{-1} b_{ng}) N^{-1} b_g}{(b_g^T N^{-1} b_g) (b_{ng}^T N^{-1} b_{ng}) - (b_g^T N^{-1} b_{ng})^2} \right)
\end{align}

These two fields are linear combinations of the two simpler fields
\begin{align}
w'_m =  N^{-1} b_g 
\end{align}
and
\begin{align}
\label{eq:wprimesigma}
w'_\sigma =  N^{-1} b_{ng} 
\end{align}
and thus contain the same amount of information. However the simpler fields have both biases non-zero.

\subsection{Fisher information with two optimal fields}

We check whether the new fields (either set) have the same total Fisher information on $\fnl$. We chose to do the calculation in the bias orthogonal basis, i.e. we define the fields $\pi_m = w^T_m \vec{\pi}$ and $\pi_\sigma = w^T_\sigma \vec{\pi}$. Our new field vector is
\begin{equation}
    \pi^{two} = \begin{pmatrix} \pi_m \\ \pi_\sigma \end{pmatrix}
\end{equation}
with biases
\begin{equation}
    b^{two}_g = \begin{pmatrix} 1 \\ 0 \end{pmatrix} \hspace{1cm} \textrm{and}  \hspace{1cm} b^{two}_{ng} = \begin{pmatrix} 0 \\ 1 \end{pmatrix}
\end{equation}
The noise covariance matrix of the new fields is given by the elements
\begin{equation}
    N_{xx'} = w_x^T N w_{x'}
\end{equation}
where $x$ stands for $m$ or $\sigma$. We find that the noise covariance is
\begin{align}
    N_{two} = \frac{1}{\det A^2} \begin{pmatrix} (n^T N^{-1} n)^2 \, g^T N^{-1} g - n^T N^{-1} n \, (n^T N^{-1} g)^2 & (n^T N^{-1} g)^3 - n^T N^{-1} g \, g^T N^{-1} g \, n^T N^{-1} n \\ (n^T N^{-1} g)^3 - n^T N^{-1} g \, g^T N^{-1} g \, n^T N^{-1} n & (g^T N^{-1} g)^2 \, n^T N^{-1} n - g^T N^{-1} g \, (n^T N^{-1} g)^2 \end{pmatrix}
\end{align}
where we have simplified the notation by writing $n=b_{ng}$ and $g=b_g$. The determinant is 
\begin{align}
\det A = g^T N^{-1} g \, n^T N^{-1} n - (n^T N^{-1} g)^2
\end{align}

As we have seen above, the Fisher information will be invariant if the bilinear forms, $b_g^T N^{-1} b_g$, $b_{ng}^T N^{-1} b_{ng}$, and $b_{ng}^TN^{-1} b_g$, remain invariant. This is indeed the case. By explicit calculation we find
\begin{align}
    b^{two,T}_g \, N^{-1}_{two} \, b^{two}_g &= b_g^T N^{-1} b_g  \\
    b^{two,T}_{ng} \, N^{-1}_{two} \, b^{two}_{ng} &= b_{ng}^T N^{-1} b_{ng}  \\
    b^{two,T}_{ng} \, N^{-1}_{two} \, b^{two}_g &= b_{ng}^T N^{-1} b_g  
\end{align}
This shows that we can compress any number of fields to two optimal fields without losing Fisher information.

\section{Linear matter field reconstruction from maximum likelihood}
\label{app:mlematter}
Assuming we have $N$ linearly-biased halo fields (or $\pi$-fields) $\delta_{h,i} = b_i\delta_m + n_i$ and $\langle n_in_j\rangle = N_{ij}$, we can write a likelihood:
\begin{equation}
  -2  \ln \mathcal{L} = (\delta_{h,i} - b_i\delta_m)(N_{ij})^{-1}(\delta_{h,j} - b_j\delta_m) 
\end{equation}
Then the maximum likelihood estimator for $\delta_m$ is obtained from the condition $\frac{\delta \ln \mathcal{L}}{\delta \delta_m}=0$.
\begin{equation}
    \frac{\delta \ln \mathcal{L}}{\delta \delta_m} = 2(\delta_{h,i} - b_i\delta_m)(N_{ij})^{-1}b_j
\end{equation}
So that
\begin{equation}
    \hat{\delta}^m = \frac{\delta_{h,i}(N_{ij})^{-1}b_j}{b_i(N_{ij})^{-1}b_j}
\end{equation}
or in matrix notation
\begin{equation}
    \hat{\delta}^m = \frac{ b^T N^{-1} \delta_{h}}{b^T N^{-1} b}
\end{equation}
Note that here $b$ is the total bias, which is only equal to $b_g$ if $\fnl=0$. We use the above result when comparing the noise of the learned matter field reconstruction $\pi_m$ with the optimal linear reconstruction $\hat{\delta}^m$.

\section{Shot noise with respect to the initial conditions}
\label{sec:shotnoise}

In the linear bias model, $\delta_h(k,z) = b(z)\delta_m(k,z)+\epsilon$, stochastic shot noise $\epsilon$ is defined with respect to the matter field $\delta_m(k,z)$ as
\begin{equation}
    N_\epsilon = \langle|\delta_h(k,z) - b(z)\delta_m(k,z)|^2\rangle.
\end{equation}
In AbacusSummit and AbacusPNG we do not have access to the late-time matter field. However, at linear scales, the matter field at late times relates to the matter field at early times by a multiplicative growth factor, so that the shot noise will be equal, whether it is defined with respect to the late-time matter field or the initial conditions (with adjusted bias). A better approach, which extends into the weakly non-linear regime, would be to evolve the initial conditions with a forward model. We test this on Quijote simulations \cite{Villaescusa-Navarro:2019bje} where we have access to the non-linear matter field as well. Figure \ref{fig:shot_lpt} shows halo shot noises at four different red shifts of Quijote simulation. The blue curves show the shot noise evaluated with respect to the scaled initial conditions and the green curve - with respect to the 2LPT-evolved initial conditions (we used 2LPT forward model from \texttt{pmwd} \cite{li2022pmwddifferentiablecosmologicalparticlemesh}). Orange curves correspond to a true shot noise with respect to a non-linear matter field. As we see, shot noise evaluated with respect to scaled initial conditions gains a slight scale dependence at intermediate scales  starting at $k=0.025\ {\rm h/Mpc}$. The effect becomes more pronounced at later red shifts. However, evolving initial densities with 2LPT gives a better (flatter) noise curve. We thus use this definition of shot noise in our analysis.

\begin{figure}[tbh!]
\centering
\includegraphics[width=1.\linewidth]{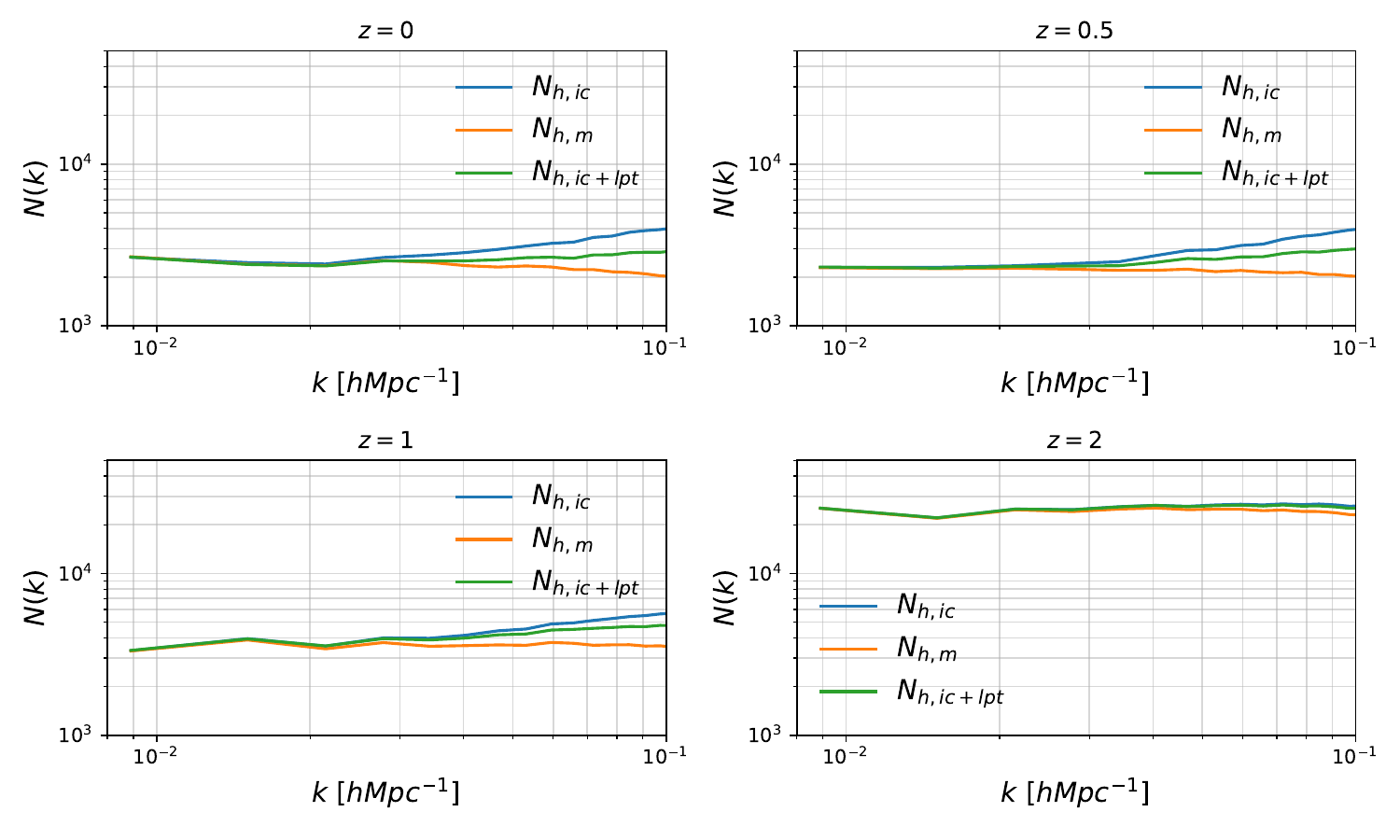}
\caption{Halo shot noise at four different red shifts of Quijote simulation, evaluated with respect to scaled initial conditions (blue), true non-linear matter field (orange), and 2LPT-evolved initial conditions (green)}
\label{fig:shot_lpt}
\end{figure}

\section{Neural Network Architecture}

\label{app:nn_arch}
For the matter reconstruction, we used the following architecture:
\begin{enumerate}
    \item $\texttt{LeakyReLU(Conv3d(n\_inp, 32, kernel\_size=3, stride=1, padding=1))}$
    \item $\texttt{LeakyReLU(Conv3d(32, 64, kernel\_size=1, stride=1, padding=0))}$
    \item $\texttt{LeakyReLU(Conv3d(64, 64, kernel\_size=1, stride=1, padding=0))}$
    \item $\texttt{LeakyReLU(Conv3d(64, 32, kernel\_size=1, stride=1, padding=0))}$
    \item $\texttt{Conv3d(32, 1, kernel\_size=1, stride=1, padding=0)}$.
\end{enumerate}
For the $\pi_\sigma(M_h)$ and $\pi_\sigma(M_h,c_h)$, the following setup is used (with the additional layer in parentheses being used only for $\pi_\sigma(M_h,c_h)$):
\begin{enumerate}
    \item $\texttt{LeakyReLU(Conv3d(n\_inp, 32, kernel\_size=1, stride=1, padding=0))}$
    \item $\texttt{LeakyReLU(Conv3d(32, 64, kernel\_size=1, stride=1, padding=0))}$
    \item ($\texttt{LeakyReLU(Conv3d(64, 64, kernel\_size=1, stride=1, padding=0))}$)
    \item $\texttt{LeakyReLU(Conv3d(64, 64, kernel\_size=1, stride=1, padding=0))}$
    \item $\texttt{Conv3d(64, 1, kernel\_size=1, stride=1, padding=0)}$
\end{enumerate}

\section{Comparison with Halo Model prediction}
\label{app:halomodel}

It is instructive to compare our analysis of AbacusSummit and AbacusPNG with the expectation from the halo model. We assume that
\begin{align}
    N_{hh,ij} = \frac{\delta_{ij}}{{n_{h,i}}} 
\end{align}
and that the non-Gaussian bias is given by the approximation
\begin{align}
\beta_i = 2 \delta_c (b_{h,i}-1).
\end{align}
We will take $\delta_c = 1.42$, as appropriate for the Sheth-Tormen halo mass function. We calculate the bias and halo density using the code \texttt{hmvec}\footnote{https://github.com/simonsobs/hmvec}. The comparison between $b^{NG}_{h,i}$, $N_{h,i}$, and $b^G_{h,i}$ calculated in halo model and estimated directly from the simulation is depicted in Fig \ref{fig:HM_nh_nh}. Halo model has a good agreement with the simulation for predicted halo number densities, as can be seen from the center plot. However, we have a slight disagreement in  $b^{NG}_{h,i}$ and $b^{G}_{h,o}$: non-Gaussian and Gaussian bias calculated with the halo model is consistently lower for higher-mass halos. 

\begin{figure}[ht]
\centering
\includegraphics[width=\linewidth ]{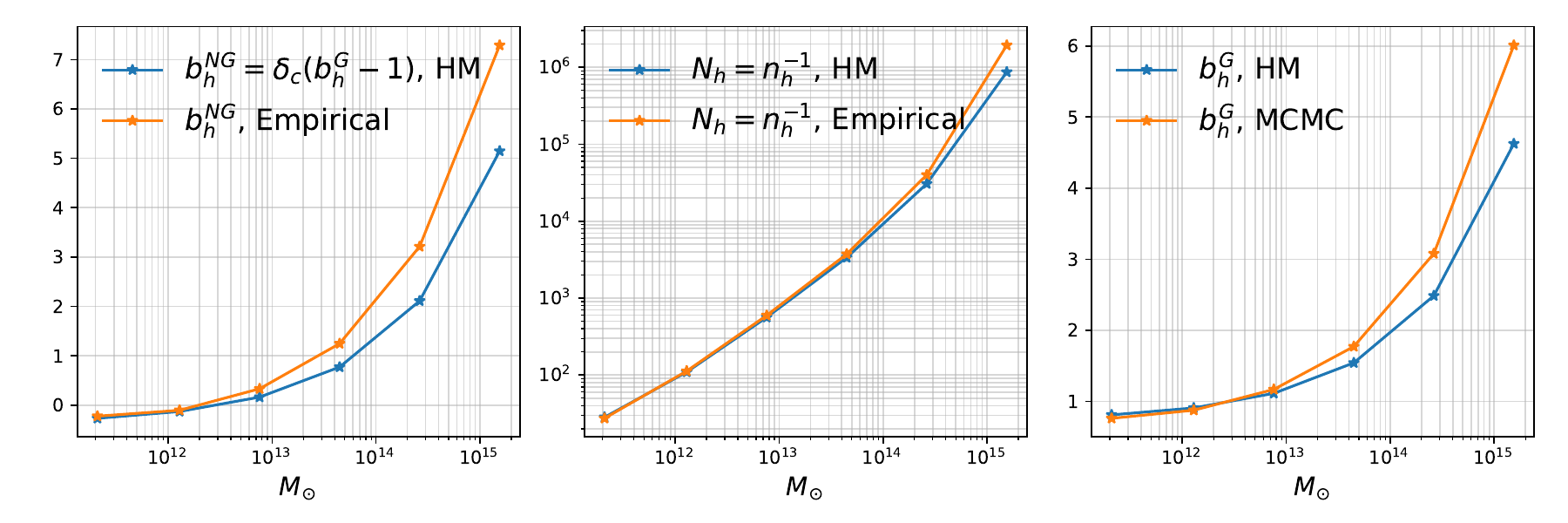}
\caption{Comparison of halo model predictions with AbacusSummit (with a halo mass cutoff equivalent to AbacusPNG). We use AbacusSummit instead of AbacusPNG so we can estimate the non-Gaussian bias from the different $\sigma_8$ values. Non-Gaussian halo bias (left), noise $N_h = n^{-1}_h$ (center), and a Gaussian bias (right) as a function of halo mass. Blue curve was obtained from halo model, orange one, left and center, corresponds to numerical estimation from AbacusSummit simulation suite. Orange curve on the right is obtained from MCMC}
\label{fig:HM_nh_nh}
\end{figure}

Figure \ref{fig:sigma_fnl_fisher_hm_mcmc} shows the comparison between the marginalized $\sigma_{f_{NL}}$ calculated according to Eq. \eqref{eq:sig_fish_marg} using the inputs from the halo model and the one obtained from the MCMC analysis. As can be seen from the plot, they are in some disagreement. The halo model predicts consistently higher $\sigma_{f_{NL}}$ for the lower mass cut-off in the range of $10^{12}-10^{14}\ M_{\odot}$ and more improvement from adding the lowest mass halos with $M<10^{12}\ M_{\odot}$.These differences can be attributed to the fact that the halo model predicts slightly different halo biases (Fig. \ref{fig:HM_nh_nh}) and assumes uncorrelated noises between bins. We also note that this analysis was performed assuming that the large-scale matter field is known and therefore less improvement is expected from adding lower-mass halos, Gaussian bias of which is close to 1.

\begin{figure}[ht]
\centering
\includegraphics[width=0.55\linewidth]{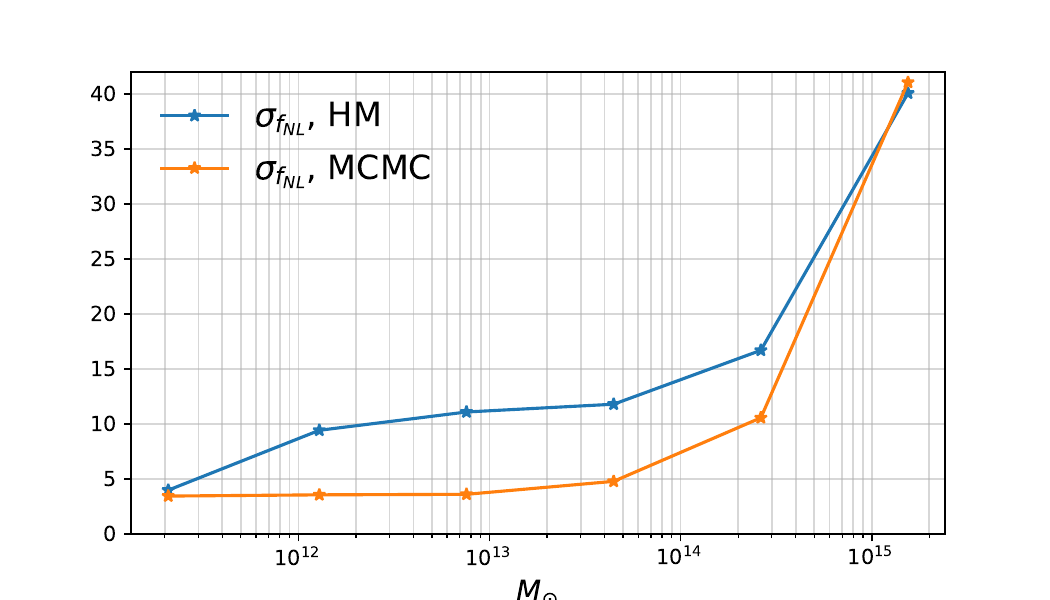}
\caption{$\sigma_{f_{NL}}$ as a function of halo mass cutoff, with mass field known. Blue curve is a fisher forecast where biases and noises were calculated in halo model. Orange curve is obtained from the MCMC analysis (same as the solid blue curve from the Figure \ref{fig:sigma_fnl_fisher_mcmc}).}
\label{fig:sigma_fnl_fisher_hm_mcmc}
\end{figure}

\end{document}